\documentclass[prd,aps,twocolumn,nofootinbib,showpacs,amsmath,superscriptaddress]{revtex4-1}

\usepackage{graphicx}
\usepackage{epstopdf}
\usepackage{bm}
\usepackage{epsfig}
\usepackage{graphics}
\usepackage{xspace}
\usepackage{amssymb}
\usepackage{amsmath}
\usepackage[dvipsnames]{xcolor}
\usepackage[utf8]{inputenc}
\usepackage[normalem]{ulem}
\usepackage{stmaryrd}

\def\OMIT#1{}

\newcommand{\df}{\mathrm{d}}

\newcommand{\nn}{\nonumber}

\newcommand{\bn}{{\bar n}}
\newcommand{\bea}{\begin{eqnarray}}
\newcommand{\eea}{\end{eqnarray}}
\newcommand{\tb}{{\bar{t}}}

\newcommand{\gsim}{\mathrel{\rlap{\lower4pt\hbox{\hskip1pt$\sim$}}\raise1pt\hbox{$>$}}}

\def\bndry{\varocircle}
\def\figeight{\circ\!\!\circ}

\newcommand{\Pythiaeight}{\textsc{Pythia}8\xspace}
\newcommand{\Pythia}{\textsc{Pythia}\xspace}
\newcommand{\Fastjet}{\textsc{FastJet}\xspace}

\newcommand{\tzcut}{\tilde z_{\rm cut}}
\newcommand{\zcut}{z_{\rm cut}}

\newcommand{\qcut}{Q_{\rm cut}}
\newcommand{\be}{\begin{equation}}
\newcommand{\ee}{\end{equation}}

\DeclareRobustCommand{\App}[1]{App.~\ref{#1}}

\DeclareRobustCommand{\Ref}[1]{Ref.~\cite{#1}}
\DeclareRobustCommand{\Refs}[1]{Refs.~\cite{#1}}

\DeclareRobustCommand{\eq}[1]{Eq.~(\ref{eq:#1})}
\DeclareRobustCommand{\eqs}[2]{Eqs.~(\ref{eq:#1}) and (\ref{eq:#2})}
\DeclareRobustCommand{\fig}[1]{Fig.~\ref{fig:#1}}
\DeclareRobustCommand{\figs}[2]{Figs.~\ref{fig:#1} and \ref{fig:#2}}
\DeclareRobustCommand{\sec}[1]{Sec.~\ref{sec:#1}}
\DeclareRobustCommand{\secs}[2]{Secs.~\ref{sec:#1} and \ref{sec:#2}}

\begin{document}
	
\title{Extracting a Short Distance Top Mass with Light Grooming}

\author{Andr\'e H. Hoang}
\affiliation{University of Vienna, Faculty of Physics, Boltzmanngasse 5, A-1090 Wien, Austria}
\affiliation{Erwin Schr\"odinger International Institute for Mathematical Physics,\\
	University of Vienna, Boltzmanngasse 9, A-1090 Wien, Austria}

\author{Sonny Mantry}
\affiliation{Department of Physics, 
                   University of North Georgia,
                   Dahlonega, GA 30597, USA}

\author{Aditya Pathak}
\affiliation{University of Vienna, Faculty of Physics, Boltzmanngasse 5, A-1090 Wien, Austria}
\affiliation{Center for Theoretical Physics, Massachusetts Institute of Technology, Cambridge, MA 02139, USA}
\author{Iain W. Stewart}
\affiliation{Center for Theoretical Physics, Massachusetts Institute of Technology, Cambridge, MA 02139, USA}

\begin{abstract}

We propose a kinematic method based on a factorization formula for precisely measuring the top quark mass $m_t$ in $pp$ collisions using boosted top jets with light soft drop grooming.  By using light grooming, which is an order of magnitude less aggressive than typical grooming, we retain a universal description of the top mass scheme and decay effects,  while still effectively removing soft contamination from the top jet. 
We give field theory results for the hadronization corrections for 
jets induced by a heavy top quark, showing they are described by a universal hadronic parameter that also appears for groomed light quark jets. 
An important phenomenological application of our results is that one can obtain $m_t$ in a short distance scheme by fitting the hadron level jet mass distributions, predicted by our factorization formula, to data or by Monte-Carlo calibration. The peaked distributions for $pp$ and $e^+e^-$ collisions are similar, up to sensitivity to underlying event which is significantly reduced by soft drop. Since soft drop implies that the $t$ and $\bar t$ jet masses each can be independently measured, the analysis enables the use of lepton+jet samples.

\end{abstract}

\pacs{12.38.Bx, 12.38.Cy, 12.39.St, 24.85.+p 
\hfill Preprint: MIT-CTP 4923, UWThPh-2017-25
}

\maketitle
 
\section{Introduction}
 
The top quark mass $m_t$ is one of the most important Standard Model (SM) parameters. It significantly affects studies of the SM vacuum stability~\cite{Buttazzo:2013uya,*Andreassen:2014gha} and the electroweak precision observables~\cite{Baak:2014ora}. The most precise top mass measurements are based on kinematic reconstruction, yielding results such as 
\begin{align}
 m_t^{\rm MC} &= 172.44(49) {\rm GeV (CMS)}~\text{\cite{Khachatryan:2015hba}} 
  \,, \\
 m_t^{\rm MC} &= 172.84(70) {\rm GeV (ATLAS)}~\text{\cite{Aaboud:2016igd}} 
  \,,\nn \\
 m_t^{\rm MC} &= 174.34(64) {\rm GeV (Tevatron)}~\text{\cite{Tevatron:2014cka}} 
  \,. \nn
\end{align} 
These measurements are based on Monte-Carlo (MC) simulations and determine the mass parameter $m_t^{\rm MC}$ of the MC generator, which depends on the  
parton shower dynamics and its interface with hadronization. Identifying these values with a Lagrangian top-mass scheme $m_t$ induces an additional ambiguity at the $0.5$--$1.0$ GeV level~\cite{Hoang:2008xm,Hoang:2014oea}. We propose a
factorization approach to remove this uncertainty in $pp\to t\bar t$ by constructing an observable that has high  kinematic sensitivity to $m_t$ and at the same time allows for hadron level predictions from QCD, employing a short distance top-mass. It can be used to extract $m_t$ from experimental data or to calibrate the parameter $m_t^{\rm MC}$ as was done for 2-Jettiness in $e^+e^-$ collisions~\cite{Butenschoen:2016lpz}. 

We consider boosted tops whose decay products are collimated in a single jet region, enabling a simultaneous and factorized theoretical description of both the top production and decay~\cite{Fleming:2007qr}.  This requires the following kinematical hierarchy
\begin{align}
\label{eq:limit}
Q \gg m_t \gg \Gamma_t \, ,
\end{align} 
where $\Gamma_t\simeq 1.4\,{\rm GeV}$ is the top width and $Q$ is the large  top jet momentum $p_J^- \equiv E_J+|\hat n_t \cdot \vec p_J|$ along the boost direction $\hat n_t$. For $pp$ collisions 
\begin{align}
 Q= p_J^- = 2p_T\cosh(\eta_J)
\end{align}
with $p_T$ and $\eta_J$ being the jet's transverse momentum and pseudo-rapidity, respectively, and where we have used the approximation in \eq{limit}. Recently an experimental analysis along these lines was carried out by CMS~\cite{Sirunyan:2017yar}. 

For $e^+e^-\to t\bar t$  a \textit{hadron level} factorization theorem for a distribution with high kinematic sensitivity to a short distance $m_t$ was derived in~\cite{Fleming:2007qr,Fleming:2007xt}. So far an analogous approach  has been missing for $pp\to t\bar t$, due to theory complications in controlling external radiation, parameters like the jet radius $R$, and soft contamination from initial state radiation and underlying event (UE) which is often modeled in MC simulations by multiple particle interactions (MPI). 

Our method relies on deriving a new factorization theorem that enables the determination of the top mass $m_t$ from the measurement of the mass $M_J$ of a top initiated fat jet of radius $R\sim 1$ with {\em light} soft drop grooming, while at the same time accounting for hadronization and underlying event effects.  The soft drop algorithm~\cite{Dasgupta:2013ihk,Larkoski:2014wba} removes peripheral soft radiation 
by comparing subsequent jet constituents $i,j$ in an 
angular-ordered cluster tree, using the Cambridge-Aachen (CA) algorithm. The grooming stops when a soft drop condition specified by fixed parameters $z_{\rm cut}$ and $\beta$  is satisfied. For $pp$ collisions the condition is
\begin{align}  \label{eq:SD}
 \frac{ {\rm min} [p_{Ti}, p_{Tj} ] }{ (p_{Ti}+p_{Tj}) } > z_{\rm cut} \Bigl( \frac{R_{ij}}{R_0} \Bigr)^\beta \,,
\end{align} 
where $R_{ij}$ is the angular distance in the rapidity-azimuth $\eta$-$\phi$ plane, $R_{ij}^2= 2(\cosh(\eta_i-\eta_j)-\cos(\phi_i-\phi_j))$, and in general $R_0$ is a parameter that is part of the definition of the soft drop algorithm. For $e^+e^-$ collisions the condition is
\begin{align} \label{eq:SDee}
 \frac{ {\rm min} [E_{i}, E_{j} ] }{ (E_{i}+E_{j}) } 
  > z_{\rm cut} \biggl( \sqrt{2}\, \frac{\sin(\theta_{ij}/2)}{\sin(R_0^{ee}/2)} \biggr)^\beta \,.
\end{align}
When Eq.~(\ref{eq:SD}) or Eq.~(\ref{eq:SDee}) is satisfied all subsequent constituents in the cluster tree are kept, thus setting a new jet radius $R_g \leq R$ for the groomed jet. 

In the limit $R_{ij}\ll 1$ with jet constituents close to the jet axis, we can also rewrite \eq{SD} in terms of energies $E_i =p_{Ti} \cosh\eta_i$ and polar angle separations $\theta_{ij}\ll 1$  as
\begin{align} \label{eq:SD2}
  \frac{ {\rm min} [E_{i}, E_{j} ] }{ (E_{i}+E_{j}) } 
  > z_{\rm cut}\,
  \Bigl( \frac{\cosh\eta_J}{R_0} \Bigr)^\beta \, \theta_{ij}^\beta
  \equiv \tilde z_{\rm cut} \, \theta_{ij}^\beta
  \,,
\end{align} 
where we have used the expansion $\cosh\eta_i = \cosh\eta_j +{\cal O}(\theta_{ij}) \>\simeq \cosh \eta_J$. 
For the factorization mode analysis in $pp$-collisions with $\theta_{ij}\ll 1$ we find it easier to use the form in \eq{SD2} in terms of $\tilde z_{\rm cut}$ defined by
\begin{align}
\tzcut = z_{\rm cut} \, \frac{\cosh^\beta (\eta_J)}{R_0^\beta} \, .
\end{align}
For $e^+e^-$-collisions, the same formula is valid but with the replacement $\tilde z_{\rm cut}^{ee} = z_{\rm cut} (\sqrt{2}\sin(R_0^{ee}/2))^{-\beta}$. 
In the original soft drop algorithm~\cite{Larkoski:2014wba} one chooses the parameter $R_0=R$, the original jet radius.
For our application, which has a sufficiently large $R$, we fix the parameter $R_0=1$. This implies that the particles kept within radius $R_g$ do not depend on the original value of $R$,  making the cross section independent of the original jet radius. 

With soft drop grooming the jet mass is defined by starting with the constituents of the jet of radius $R$ and summing only over those constituents that remain in the groomed jet, ${\cal J}_{sd}$: 
\begin{align} \label{eq:mJ}
M_J^2 = \biggl(\: \sum_{i\in {\cal J}_{sd}} p_i^\mu \biggr)^2 \,. 
\end{align}  
The measurement of $M_J$ is performed on a hadronically decaying top quark, and hence can be applied to both the all jets and lepton+jets $t\bar t$ samples. 
This grooming retains strong kinematic sensitivity to $m_t$ as in direct reconstruction methods, removes contamination from other parts of the collision, and allows for a factorization based description as demonstrated for massless jets~\cite{Frye:2016aiz}. Monte-Carlo studies of top quarks with soft drop have also shown to have reduced tuning dependence~\cite{Andreassen:2017ugs}.

We use the Soft-Collinear Effective Theory (SCET)~\cite{Bauer:2000ew,*Bauer:2000yr,*Bauer:2001ct,*Bauer:2001yt,*Bauer:2002nz} to derive a factorization formula for the groomed top jet mass distribution  in the peak region, $(M_J^2 -m_t^2)/m_t \sim \Gamma_t \ll m_t$, for boosted top quark pair production,  where the hierarchy in \eq{limit} applies.  The grooming parameters are chosen in the light grooming region, such that significant contamination is removed from the top jet while retaining the top decay products and leaving the ultra-collinear (UC) radiation associated with the top quark unaffected. This allows for a simple treatment of top decay products and a clear interpretation of the short distance top mass scheme. For a fat top quark initiated jet  the light grooming region is identified by the conditions:
\begin{subequations}\label{eq:zcutcons0}
\begin{align} 
\label{eq:zcutcons}
& \zcut \lesssim \>\frac{\Gamma_t}{ h^{2 + \beta}  m_t} \Big(\frac{p_T}{m_t }\Big)^\beta 
,\,
 \\
\label{eq:zcutcons2}
& \zcut^{\frac{1}{2+\beta}} \gg 
\frac{1}{2} 
\bigg( \frac{\Gamma_t}{m_t}\frac{m_t^2}{ p_T^2} \bigg)^{\frac{1}{2+\beta}}
\,.
\end{align}
\end{subequations}
The dimensionless function $h$, defined below in \eq{hdef}, is related to the angles of the top decay products with respect to the jet axis. For the $Q$ values of interest it has an average value around $\langle h\rangle \sim 2$.

 The first constraint in \eq{zcutcons} enables a simple treatment of the top-decay products. It also implies that $\zcut \ll \Gamma_t/m_t (p_T/m_t)^\beta$, the necessary condition to ensure that the ultra-collinear radiation associated with the top quark, is unaffected by the grooming. This condition is significantly stronger than the condition, $\zcut \ll (p_T/m_t)^\beta $, needed to retain the high energetic contributions of the top decay products, and guarantees that there will be a clear peak directly depending on the top mass value.   In the factorization theorem discussed below, the constraint of \eq{zcutcons} on the grooming parameters allows for a clear interpretation of the top mass scheme, determined by the  heavy quark jet function~\cite{Fleming:2007qr} which describes the inclusive dynamics of the ultra-collinear  radiation and the evolution and decay of the top quark in the peak region.  
 
 The second constraint in \eq{zcutcons2} ensures that wide angle soft radiation is groomed away,  isolating the jet and removing the majority of soft contamination. The conditions in \eq{zcutcons0} follow directly from a SCET analysis of the relevant modes in the presence of the soft drop and jet mass constraints, as we discuss in \secs{topmodes}{topkinematics}.

\begin{figure}[t!]
	\includegraphics[width=0.45\textwidth]{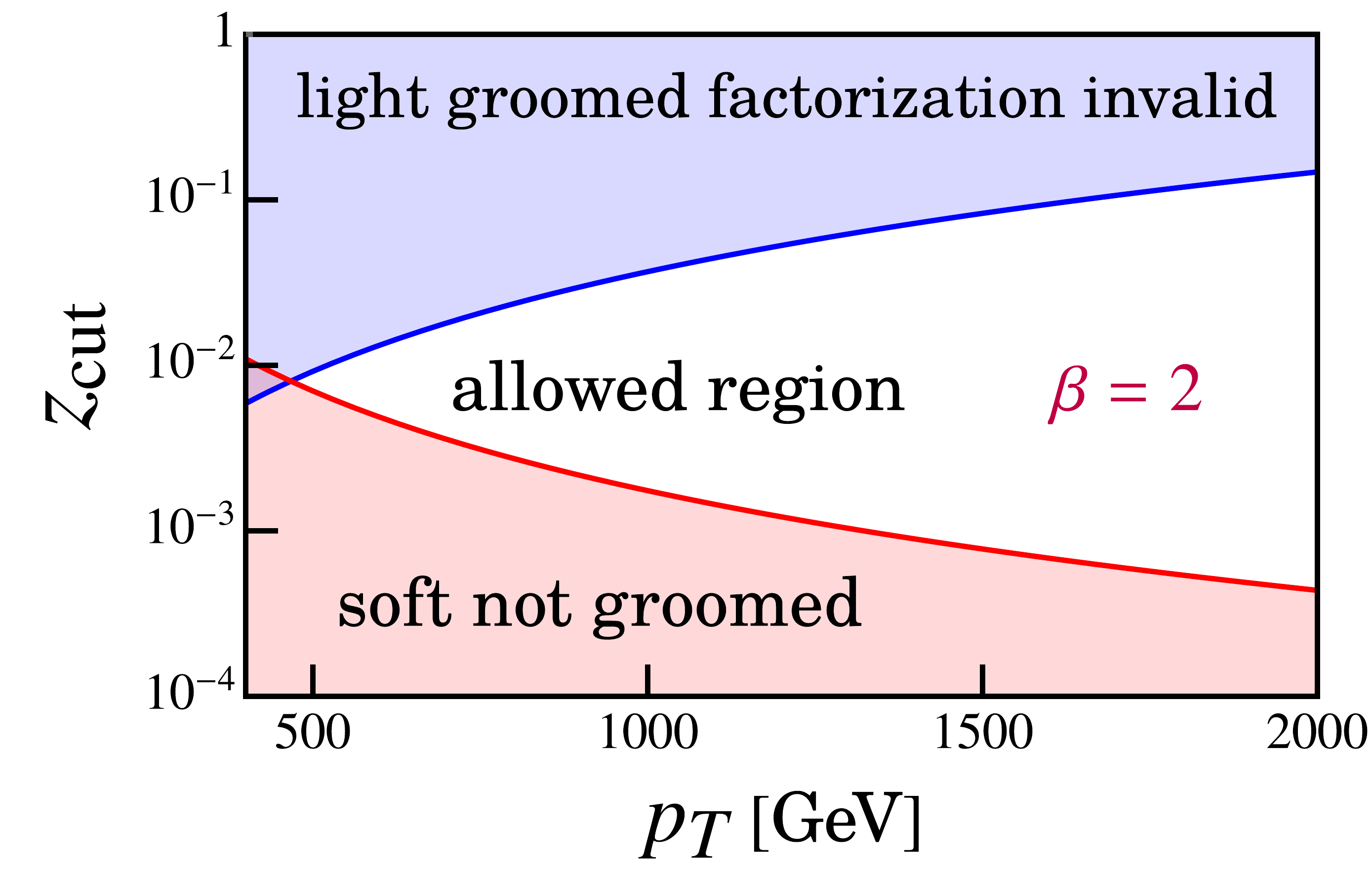}\\
	\vspace{-0.2cm}
	 
	\caption{Allowed values of $z_{\rm cut}$ which are strong enough to isolate the jet from contaminating radiation (above red band), but not so strong as to invalidate the factorization formulae we derive (below blue band).  
		\label{fig:expansion} 
	}
	\vspace{-0.2cm}
\end{figure} 

The allowed $z_{\rm cut}$ region satisfying \eq{zcutcons0} is shown as a function of $p_T$ in Fig.~\ref{fig:expansion} for a jet with $\eta_J=0$ and $\beta=2$. The upper blue line is obtained by replacing ``$\gtrsim$" by an equality in \eq{zcutcons} and the lower red line by replacing the ``$a\gg b$'' by ``$a=3 b$'' in \eq{zcutcons2}.  Using $p_T=750\,{\rm GeV}$
and setting $h=2$, the constraints in \eq{zcutcons0} become $z_{\rm cut} \lesssim 0.01 $ and $z_{\rm cut}^{1/4} \gg .066$ which is satisfied by $z_{\rm cut}\simeq 0.01$.  This light grooming is an order of magnitude smaller than typically used for many analyses at the LHC, but as we will see, is still very effective for $m_t$ measurements. 
 We take $\beta=2$ as our default choice.
Analyses for $\beta<2$ are also viable, but the allowed region is more constrained, indicating that for experimentally accessible $p_T$ values the expansions used to derive the factorization formula are less convergent. It is known that soft drop reduces pileup corrections, and although we will not include these effects in the study done here, it would be worth doing so in the future. We comment further on what such a pileup study should address in our conclusions.

In this paper we present the key aspects of the factorized jet mass cross-section with a focus on results with a next-to-leading-logarithmic (NLL) resummation of large logarithms in the partonic cross section, including hadronization corrections whose perturbative coefficients are calculated at LL accuracy. In Ref.~\cite{Butenschoen:2016lpz,Fleming:2007qr,Fleming:2007xt} it was shown that nonperturbative hadronization corrections play an important role for measurements of the top mass from boosted top jets in $e^+e^-\to t\bar t$ for achieving a precision below 1 GeV. This remains true for $pp\to t\bar t$.  We will review results from \Ref{Hoang:2019ceu} that provide a description of hadronization corrections derived from field theory for soft drop groomed jets that are initiated by either massless quarks or gluons. Here we develop formalism for treating jets initiated by heavy and unstable top quarks.  We show that although the leading hadronization effects are influenced by the presence of soft drop, they are independent of the jet  $p_T$, jet rapidity $\eta_J$, and the soft drop parameter $\zcut$. This universality is important for fits to $\alpha_s$ for massless groomed jets~\cite{Baron:2018nfz}.  Interestingly, for groomed top jets we will show that the hadronization corrections are also independent of $\beta$ and depend on the same non-perturbative parameter $\Omega_{1q}^{\figeight}$ that appears for massless quark initiated groomed jets.  This is in contrast to massless quark and gluon initiated groomed jets where two additional $\beta$-independent non-perturbative parameters are required to describe hadronization effects. We also discuss in detail the most important phenomenological results for the measurement of the top mass.  

The outline of the paper is as follows. In \sec{massless} we discuss the hadron level factorization for groomed massless jets~\cite{Hoang:2019ceu} in order to summarize the influence of  non-perturbative hadronization on the groomed jet-mass. In \sec{massive} we then turn to discuss the factorization theorem for jets initiated by the massive boosted top quark, which requires a careful treatment of the top decay products in the presence of soft drop grooming. Here we also describe our treatment of leading hadronization corrections and show that  with certain systematic approximations the top jet-mass cross section has a leading hadronization parameter that is the same as the one that appears in the massless case examined in \Ref{Hoang:2019ceu}. In \sec{results} we present results for the factorization theorem and a first calibration study for the top-mass in $pp$ collisions made by comparison to \Pythiaeight.  We give a more detailed  analysis of both the factorization and Monte-Carlo results in Ref.~\cite{Hoang:2018SDlong}.

\section{Hadron Level Factorization for Light Quark and Gluon Jets} \label{sec:massless}

\subsection{Effective Theory Modes}

In this section we review the partonic massless soft drop factorization theorem derived in~\cite{Frye:2016okc,Frye:2016aiz}, and its extension to hadron level given in ~\cite{Hoang:2019ceu}. The latter was achieved by incorporating non-perturbative hadronization parameters based on a field theory operator expansion, 
which account for the dominant final state hadronization effects. 

Consider the groomed jet mass measurement on jets initiated by light quarks or gluons.  Note that in this section we do not require light soft drop grooming and work in the same limit as in Refs.~\cite{Larkoski:2014wba,Frye:2016aiz}
\begin{align} \label{eq:sdlimit}
 \frac{m_J^2}{Q^2} \ll \tzcut \ll 1 \,.
\end{align}
Here small  $m_J$ denotes the jet mass for jets initiated by light quarks and gluons in order to distinguish it from the top jet mass $M_J$, despite the fact that the definition of $m_J$ is the same as in \eq{mJ}.
The use of \eq{sdlimit} yields a fully factorized description of the groomed jet mass cross section in the SCET framework. The physics of the perturbative radiation in this  jet mass region also plays a role for boosted tops in the peak region, as discussed in \sec{massive}. The parton level factorization formula for the groomed jet mass measurement on jets initiated by light quarks or gluons \cite{Frye:2016okc,Frye:2016aiz} reads\footnote{The notation used here follows \Ref{Hoang:2019ceu} and differs slightly from \Refs{Frye:2016aiz}.}
\begin{align} \label{eq:masslessFact0}
 &\frac{d^2 \hat \sigma}{d m_J^2d\Phi_J}= \sum_{\kappa={q,g}}  N_\kappa(\Phi_J, R,\zcut, \beta,\mu)
 \\
 & \qquad \times\int\!\! d\ell^+ \, J_\kappa \big(m_J^2 - Q\ell^+ ,\,  \mu \big) \qcut^{\frac{1}{1+\beta}} \: S_c^\kappa \Big[ \ell^+ \qcut^{\frac{1}{1+\beta}},\beta,\mu\Big] 
 \,.\nn
 \end{align}
Here the index $\kappa$ denotes the partonic channel ($\kappa=q,g$). The normalization factor, $N_\kappa(\Phi_J,R, \zcut, \beta,\mu)$, encodes the underlying hard process and the PDFs. It also accounts for the soft radiation that is groomed away by soft drop, and hence depends on the soft drop parameters $\zcut$ and $\beta$, in addition to the jet kinematic variables denoted by $\Phi_J=\{p_T,\eta_J\}$ and the jet radius R. It also determines the fractional contribution from each partonic channel.  The factorization formula also involves the inclusive jet function $J_\kappa$, which describes the dynamics of the collinear radiation, and the collinear-soft function $S_c$, which describes the dynamics of soft radiation affected by soft drop, which has an impact on the jet mass measurement.

Each of the functions in \eq{masslessFact0} involves matrix elements of a single type of quark and gluon fields, corresponding to distinct  modes in the effective theory. The modes are distinguished by their momentum scaling, which we briefly review. The collinear radiation contained in $J_\kappa$ has the scaling
\begin{align}
p_C^\mu \sim \Bigl( \frac{m_J^2}{Q}, Q, m_J \Bigr) \,,
\end{align}  
where we use light-cone components 
\begin{align}
p^\mu \sim (p^+,p^-,p_\perp)\equiv  (n_J\cdot p,\bar n_J\cdot p,p_\perp) \,,
\end{align}
relative to the jet axis $\hat n_J$, with $n_J^\mu=(1,\hat n_J)$ and $\bar n_J^\mu=(1,-\hat n_J)$. Since these collinear modes lie at the smallest angles $\theta \sim 2m_J/Q$ and have the highest energies $E \sim Q/2$, they automatically pass soft drop so that $J_\kappa$ is independent of the grooming parameters.

The softer modes at wider angles may or may not pass the soft drop. Global soft modes which do not affect the jet mass spectrum have scaling $p_{S}^\mu \sim Q z_{\rm cut}$. The soft modes that effect the jet mass spectrum are boosted along the jet's direction, and are called the collinear-soft modes. They are dominated by the radiation that lies at the widest angle and has the smallest energy needed to both pass the soft drop test and contribute to $m_J$~\cite{Frye:2016aiz}. They follow the momentum scaling
\begin{align} \label{eq:pCS}
p_{cs}^\mu &\sim  \frac{m_J^2}{Q\zeta_0} 
\Bigl( \zeta_0 ,\frac{1}{\zeta_0} , 1 \Bigr)
\,,
\qquad \zeta_0 \equiv \Bigl( \frac{m_J^2}{Q \qcut} \Bigr)^{\frac{1}{2+\beta}}
\,,
\end{align}
which depends on a single combination of $\zcut$ and $Q$ via the soft drop modified hard scale $\qcut$ defined as
\begin{align}
\label{eq:qcut}
Q_{\rm cut} \equiv 2^\beta  Q\,  \tilde z_{\rm cut} 
= \Big(\frac{2\cosh\eta_J}{R_0}\Big)^\beta Q z_{\rm cut} \,.
\end{align}
The collinear-soft modes lie at an angle $\theta$ that is given by
\begin{align} \label{eq:thetaCSdef}
\theta \sim  2 \sqrt{\frac{p_{cs}^+}{p_{cs}^-}} \sim 2 \, \zeta_0 \, .
\end{align}
The modes maintain a collinear scaling between their momentum components since $\zeta_0 \ll 1$ precisely because $m_J^2/Q^2 \ll \zcut$.

The collinear soft function has the following operator definition:
\begin{align} \label{eq:SCdefn}
& S_c^\kappa\Bigl[\ell^+\, Q_{\rm cut}^{\frac{1}{1+\beta}}, \beta \Bigr]
\\
&\ \ 
\equiv \frac{Q_{\rm cut}^{\frac{-1}{1+\beta}}}{n_\kappa} \:
{\rm tr} \bigl\langle 0 \big|\bar  T X_{n\kappa}^\dagger V_{n\kappa} \delta \bigl(\ell^+- \overline\Theta_{\rm sd}\, \hat p_{cs}^+\bigr) T V_{\kappa n}^\dagger X_{n\kappa} 
\big| 0 \bigr\rangle 
. \nn
\end{align}
Note that in this notation $S_c$ has mass dimensions $(-2-\beta)/(1+\beta)$ and depends on the single combination $\ell^+ \qcut ^{\frac{1}{1+\beta}}$. Here $n_q=N_c$ and $n_g=N_c^2-1$ normalize the color trace and $N_c=3$ is the number of colors. $\overline\Theta_{\rm sd}$ is a soft drop measurement function that selects the collinear-soft particles that pass soft drop. The Wilson lines $V_{n\kappa}=V_{n\kappa}[\bn\cdot A_{cs}]$ and $X_{n\kappa}=X_{n\kappa}[n\cdot A_{cs}]$ consist of collinear-soft fields in  the fundamental representation ($\kappa=q$) or the adjoint representation ($\kappa=g$).

In \fig{modesmassless} we show  the perturbative modes that enter the factorization formula in \eq{masslessFact0} in the $\ln (z^{-1})$--$\ln (\theta^{-1})$ plane. The location of the perturbative collinear (C), collinear-soft (CS), and global soft modes (S) in \fig{modesmassless} is determined by the jet mass measurement $m_J$ and the soft drop criteria $z \gtrsim \zcut \theta^\beta$. Any emission from the jet-initiating parton that yields an observed groomed jet mass $m_J^2 = Q p_J^+$ must lie on the blue line and outside the shaded region that is groomed away by soft drop. This groomed region is delineated by the slanted orange line for $z\simeq z_{\rm cut} \theta^\beta$ labeled by ``$\text{slope} =\beta$'' and the angle of the widest emission that first passes soft drop and stops further grooming. We define $\theta_{cs}$ as the angle relative to the jet axis of the CS radiation that stops soft drop. This angle then determines the groomed jet radius $R_g$ as shown in \fig{modesmassless}. The effects of recoil of the collinear subjet only appear at subleading order in the power expansion. Finally, the global soft modes (S) ensure renormalization consistency and enter into the calculation of $N_{\kappa}$. In the region of large jet mass, corresponding to moving the blue line downwards, the collinear-soft and the global soft modes merge leading to a transition into the ungroomed region where soft drop is not active anymore. This happens for jet masses satisfying $m_J^2/p_T^2 \gtrsim \zcut$, which is the ungroomed resummation region, and requires further merging of soft components of the partonic factorization formula in \eq{masslessFact0}.

\subsection{Nonperturbative Modes}

\begin{figure}[t!]
	\includegraphics[width=0.45\textwidth]{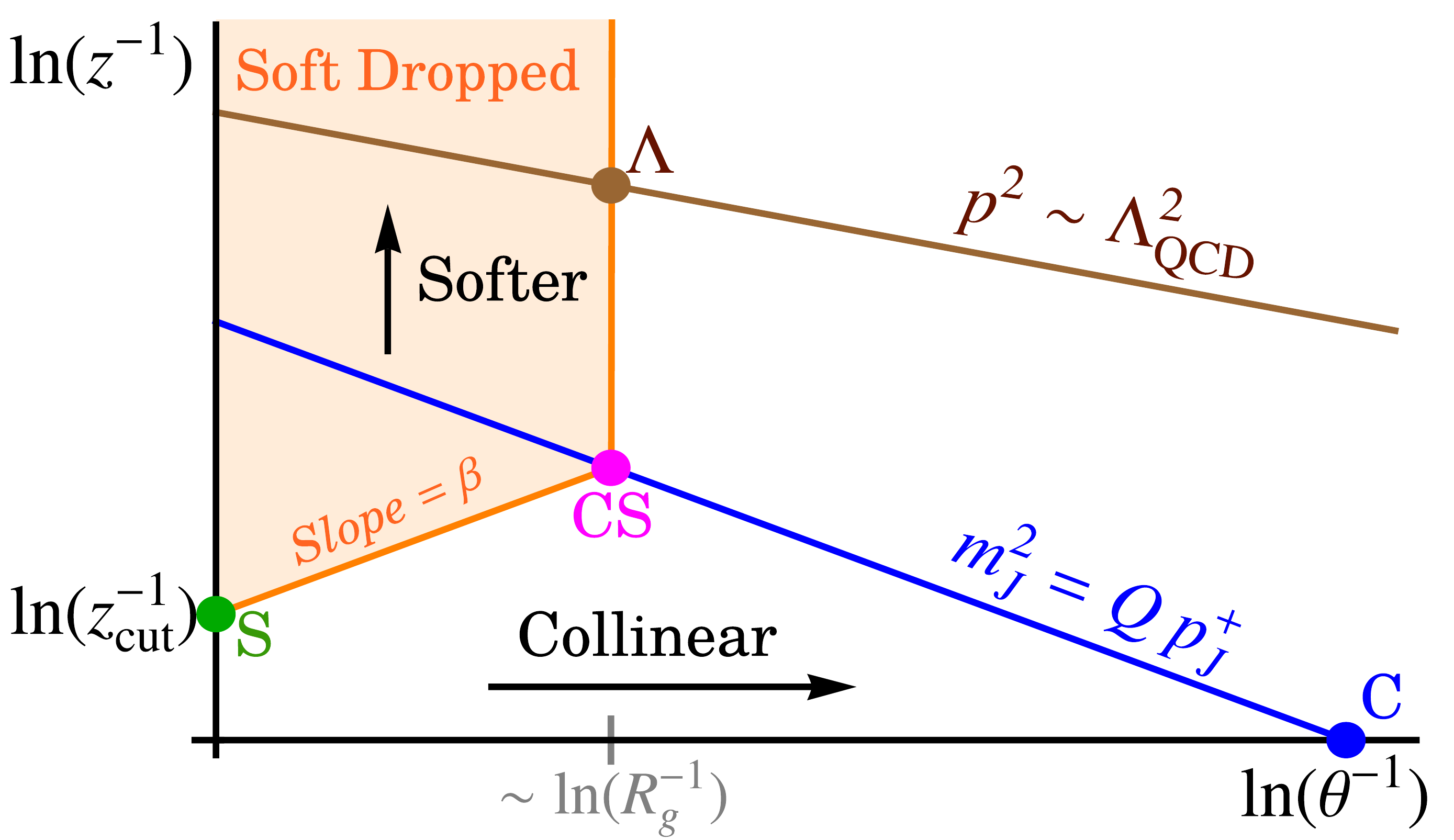}\\
	\vspace{-0.2cm}
	 
	\caption{Picture of modes appearing in the factorization formula for jets initiated from massless quarks or gluons. Here $z=2E/Q$ is the energy fraction for energy $E$, and $\theta$ is the polar angle relative to the jet-axis. 
		\label{fig:modesmassless} 
	}
	\vspace{-0.2cm}
\end{figure} 
Next we consider the extension of  \eq{masslessFact0} to account for hadronization, reviewing results from \Ref{Hoang:2019ceu}. The brown line $p^2\sim \Lambda_{\rm QCD}^2$ in \fig{modesmassless} indicates the modes that are responsible for hadronization and span all angles. In the region of the jet mass spectrum, where the CS mode that stops soft drop is perturbative, the dominant nonperturbative mode $\Lambda$ is determined by the intersection of lines corresponding to $\theta = \theta_{cs}$ and  $p^2\sim \Lambda_{\rm QCD}^2$. The $\Lambda$ mode has the same parametric boost as the CS mode, with the momentum scaling given by
\begin{align} \label{eq:pL} 
p_\Lambda^\mu &\sim \Lambda_{\rm QCD} \Bigl( \zeta_0 ,\frac{1}{\zeta_0} , 1 \Bigr) \,.
\end{align}
The $\Lambda$ mode has the largest allowed $p^+$ component among all the NP modes,  and hence yields the leading nonperturbative contribution to the jet mass measurement $m_J^2 = Q p^+$. Note that this implies that in the region where the CS modes are perturbative they also determine the boost for the $\Lambda$ mode in \eq{pL}. This is satisfied when $p_{cs}^+ \gg p_{\Lambda}^+$, such that the two modes are hierarchically separated in their $z$ values, referring to their vertical separation in \fig{modesmassless}. This corresponds to the jet mass region satisfying
\begin{align}  \label{eq:sdoe}
\frac{Q \Lambda_{\rm QCD}}{m_J^2}
   \Bigl( \frac{m_J^2}{Q Q_{\rm cut}}\Bigr)^{\frac{1}{2+\beta}}  
   \ll 1 
  \, , 
\end{align}
which puts a lower bound on $m_J$.
We refer to this region of the jet mass spectrum as the soft drop operator expansion (SDOE) region. 

On the other hand, in the region of smaller jet masses the CS and $\Lambda$ modes merge together when $p_{cs}^+\sim p_\Lambda^+$. This happens in the jet mass region
\begin{align} \label{eq:mJNP}
m_J^2 \sim Q\Lambda_{\rm QCD} (\Lambda_{\rm QCD}/Q_{\rm cut})^{\frac{1}{1+\beta}} \, .
\end{align} 
In other words, the CS mode for these jet masses is nonperturbative and the contribution from the $\Lambda$ modes is no longer power suppressed. This region is referred to as the Soft Drop Nonperturbative (SDNP) region.

\subsection{Nonperturbative Power Corrections}\label{sec:masslessfact}

In what follows, we limit our discussion to reviewing results for the SDOE region from~\cite{Hoang:2019ceu}, where a factorized description of the dominant nonperturbative corrections of the hadron level cross section is possible in terms of a small number of hadronic parameters. We will see later that it is this region that is also relevant for groomed top jets.

\begin{figure}[t!]
	\includegraphics[width=0.3\textwidth]{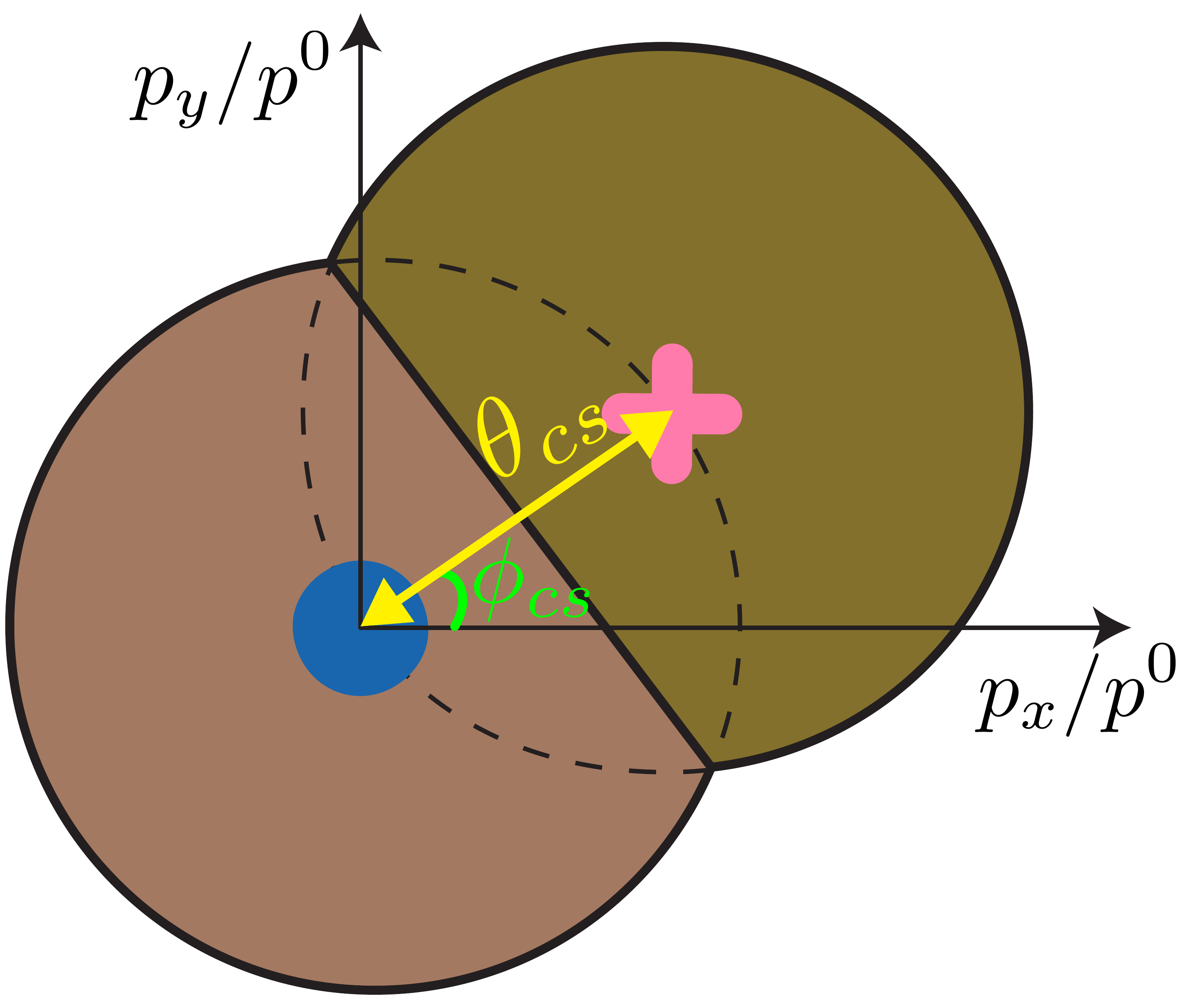}\\
	\vspace{-0.2cm}
	 
	\caption{The catchment area of the non-perturbative modes kept by the soft drop factorization formula at LL, pictured from above looking down the jet axis. These modes are clustered with either the collinear subjet located on the jet axis (blue dot) or the collinear-soft subjet (pink cross), as indicated by the shaded brown regions. The overlapping circles both have radius $\theta_{cs}$.
		\label{fig:NPfig8} 
	}
	\vspace{-0.2cm}
\end{figure}

From \eq{pL} and \fig{modesmassless} we note that the power corrections to the groomed jet mass are intricately tied to the perturbative branching history which determines the soft drop stopping angle $\theta_{\rm stop} \simeq 2\zeta_0$. This is quite unlike the case of ungroomed event shapes where the power correction is described by a single parameter that is not modified by the resummation of logarithms between perturbative scales, and does not depend on the intrinsic geometric details of the event. Here the structure of the power corrections depends on the opening angle of the stopping pair which sets the catchment area for the nonperturbative (NP) particles. In general, due to the CA clustering the region of phase space where NP particles get clustered with the groomed jet can be quite complex and intricately tied to the perturbative branching history. However, at LL accuracy, we can assume strong angular ordering of the perturbative emissions. As a result, the  catchment area in momentum space for the $\Lambda$ modes is given by a pair of overlapping cones, as shown by the brown shaded region in \fig{NPfig8}.
The LL approximation for determining this region ensures that the subsequent perturbative emissions that are kept by the soft drop lie at much smaller angles and hence do not influence the clustering of NP modes with the collinear and collinear-soft subjets.
Each cone is centered on one of the pair of subjets that stops soft drop, and the conic sections correspond to the non-perturbative radiation collected by each of these subjets.  Since the polar angle of the CS subjet relative to the collinear top jet axis is $\theta_{cs}\ll 1$, for the axis scaled in the manner shown, the two circles simply have radius $\theta_{cs}$.  This catchment area determines the amount of nonperturbative radiation that contributes to the jet mass measurement for the ``shift'' power correction. The essential point is that the catchment area of the captured nonperturbative radiation is tied to the soft drop stopping angle, $\theta_{cs}$, which is predominantly determined by the perturbative dynamics~\cite{Hoang:2019ceu}.

\begin{figure}[t!]
	\includegraphics[width=0.3\textwidth]{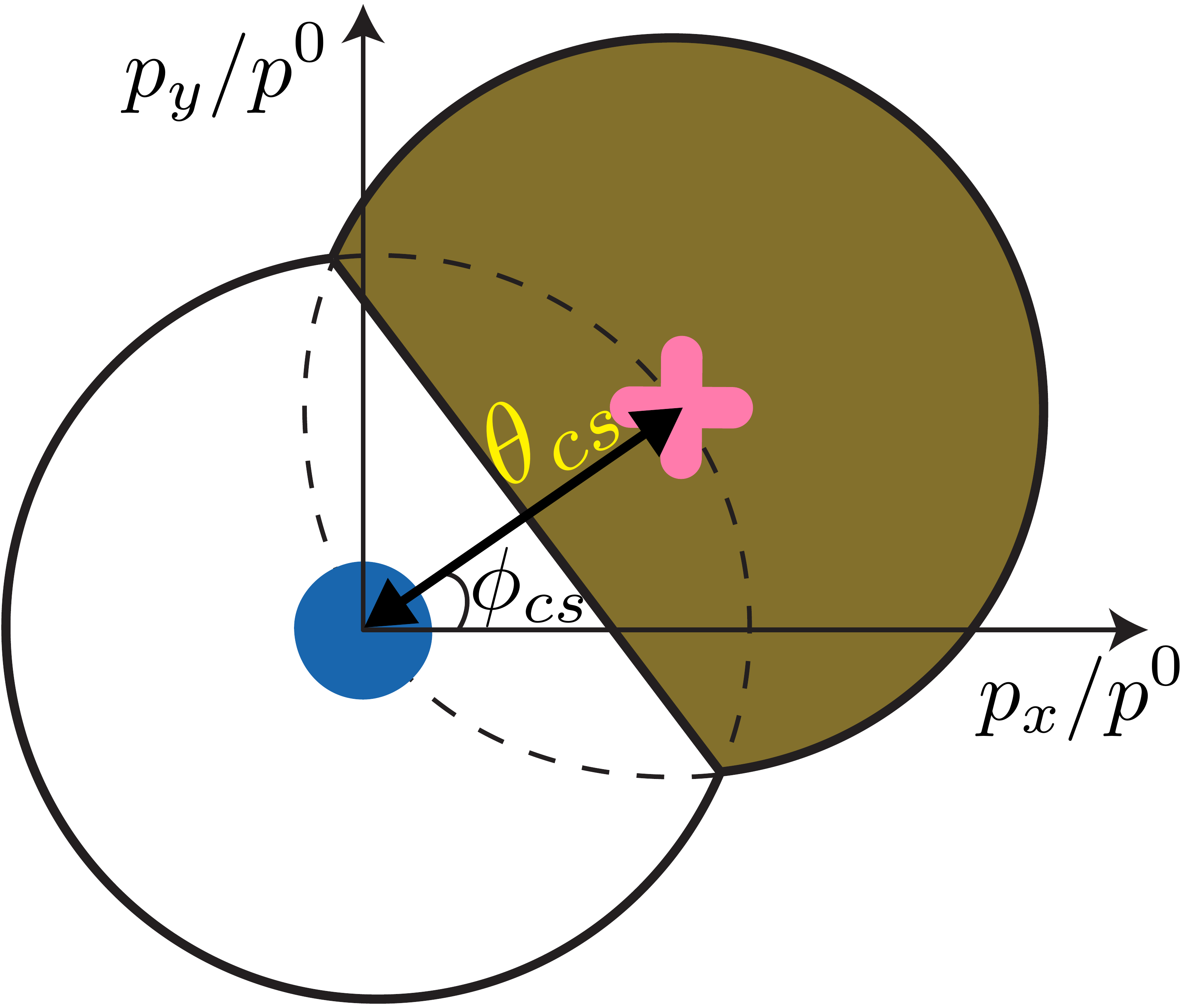}\\
	\vspace{-0.2cm}
	 
	\caption{The region in momentum space that is relevant for the nonperturbative modes that influence the soft drop comparison test for the collinear-soft subjet. The plot is pictured from above looking down the jet axis. These modes are clustered with the collinear-soft subjet (pink cross), but not with the collinear subjet (blue dot). 
		\label{fig:NPfig8bndry} 
	}
	\vspace{-0.2cm}
\end{figure}

A second effect of the nonperturbative modes is their influence on the soft drop comparison condition, which we refer to as the ``boundary" power correction. As an example scenario, consider the  \textit{i}-th collinear-soft subjet tested for soft drop with total momentum $p_i^\mu + q_i^\mu$, where $p_i^\mu$ and $q_i^\mu$ denote the momentum contributions of the perturbative and nonperturbative modes in the subjet, respectively. The soft drop condition then reads
\begin{align}
\label{eq:sdthetapq}
\overline \Theta_{\rm sd}^{\, p_i+q_i}\! &= \Theta \bigg ( \frac{p_i^- + q_i^-}{Q} - \tzcut \, \Big( \frac{2 |\vec p_{i,\perp} + \vec q_{i,\perp}| }{p_i^- + q_i^-} \Big)^\beta \bigg) \nn \\
&=\overline \Theta_{\rm sd}^{\, p_i}  + \delta \big ( z_{p_i} - \tzcut \theta_{i}^\beta \big)\nn \\
&\qquad \times  \frac{q_i^- }{Q} \, \bigg (\, (1 + \beta) - \beta \, \frac{\theta_{q_i}}{\theta_{i}} \, \cos(\Delta \phi) \bigg) \, ,
\end{align}
with $\overline \Theta_{\rm sd}^{\, p_i}$ denoting the soft drop condition applied to the perturbative momentum  $p_i$ alone and $\Delta \phi = \phi_{q_i} - \phi_{p_i}$ is the relative azimuthal angle between $\vec p_i$ and $\vec q_i$.  The opposite scenario, where the subjet loses NP momentum, corresponds to replacing $q_i^-$ with $-q_i^-$ in \eq{sdthetapq}. Thus the soft subjets that marginally fail or pass the soft drop test will be affected by the clustered NP modes. The contribution to this power correction from soft drop failing subjets enters beyond LL accuracy, and hence we only need to consider this effect on the final soft drop stopping collinear-soft subjet~\cite{Hoang:2019ceu}. The relevant region of phase space at LL corresponds to all the NP particles that get clustered with (or are lost from) the stopping collinear-soft  subjet, and is displayed as the brown shaded region in \fig{NPfig8bndry}. Here the two circles in the boosted limit again have the radius $\theta_{cs}$.

In the region of jet mass spectrum illustrated in \fig{modesmassless}, where the perturbative modes are well separated from the NP modes, the key ingredient that determines the size of both of these power corrections is the angle $\theta_{cs}$ of the perturbative collinear-soft subjet relative to jet axis. The projection operator $\overline \Theta_{\rm NP}^{\, \figeight}$ that selects NP radiation responsible for the shift correction, shown in \fig{NPfig8}, is
\begin{align}
\label{eq:thetashift}
\overline \Theta_{\rm NP}^{\, \figeight}(  p_{\Lambda }^\mu, \, &\theta_{cs}, \phi_{cs})  = \Theta\bigg(| \Delta \phi| -  \frac{\pi}{3} \bigg)\Theta\bigg(1 -\frac{ \theta_{\Lambda}}{\theta_{cs}} \bigg) \\
&+ \Theta\bigg(\frac{\pi}{3} - | \Delta \phi| \bigg)\Theta \bigg(2 \cos (\Delta \phi) - \frac{\theta_{\Lambda}}{\theta_{cs}}\bigg)\, . \nn
\end{align}
Similarly, the projection operator $\overline \Theta_{\rm NP}^{\, \bndry}$ that selects NP radiation responsible for the boundary correction, shown in \fig{NPfig8bndry}, is
\begin{align}
\label{eq:thetabndry}
\overline \Theta_{\rm NP}^{\, \bndry} (p^\mu_{\Lambda}, \, \theta_{cs},\, \phi_{cs}) &= \Theta\bigg(\frac{\pi}{3} - | \Delta \phi| \bigg)\Theta \bigg(\frac{\theta_{\Lambda}}{\theta_{cs}} - \frac{1}{2 \cos (\Delta \phi)}\bigg) \nn  \\
&\qquad\times \Theta \bigg(2 \cos (\Delta \phi) - \frac{\theta_{\Lambda}}{\theta_{cs}}\bigg) \, ,
\end{align}
where  $\Delta \phi = \phi_\Lambda - \phi_{cs}$. The projection operators, $\overline \Theta_{\rm NP}^{\, \figeight}$ and $\overline \Theta_{\rm NP}^{\, \bndry} $, are equal to 1 if the NP momentum $p_\Lambda$ lies in the respective shaded regions shown in \fig{NPfig8} or  \fig{NPfig8bndry} respectively,  and zero otherwise. 
Noting that only  the ratio $\theta_\Lambda/\theta_{cs}$ appears in \eqs{thetashift}{thetabndry}, we can further simplify the expressions by expressing the NP momenta in a boosted and rotated frame determined by the CS subjet~\cite{Hoang:2019ceu}:
\begin{align} \label{eq:rescaleNP}
p_{\Lambda }^+  &= \frac{\theta_{cs}}{2} \: k^+ 
\,, \qquad  
p_{\Lambda }^\perp = k^\perp
\,, \qquad 
p_{\Lambda }^- = \frac{2}{\theta_{cs}}\: k^-
\,,\nn\\
 \phi_\Lambda &= \phi_k + \phi_{cs} \, .
\end{align}
As a result we find that the rescaled momenta scale homogeneously:
$k^+ \sim k^- \sim k^\perp \sim \Lambda_{\rm QCD}$, where in terms of the new variables we have $\theta_{\Lambda}/\theta_{cs} = \frac{k_\perp}{k^-}$  and $\Delta \phi = \phi_k$.  The projection  $\overline \Theta_{\rm NP}^{\, \figeight}$ now solely depends on $k^\mu$ variables:
\begin{align}
\overline \Theta_{\rm NP}^{\, \figeight} \Big(\frac{k_\perp}{k^-},& \,  1, \,  \phi_k \Big)  \equiv \Theta\bigg(| \phi_k| -  \frac{\pi}{3} \bigg)\Theta\bigg(1 -\frac{ k_\perp}{k^-}\bigg) \\
 &+ \Theta\bigg(\frac{\pi}{3} - | \phi_k| \bigg)\Theta \bigg(2 \cos ( \phi_k) - \frac{ k_\perp}{k^-}\bigg) \nn \,,
\end{align}
where the second the argument being $1$ emphasizes that in the rescaled coordinates the cones in \fig{NPfig8} now have radius 1.
Thus the contribution of an NP particle with momentum $q^\mu$ to the jet mass is given by 
\begin{align}
\label{eq:shift}
Q\, q^+ = (\theta_{cs}/2) Q \, k^+ 
\end{align}
when 
$\overline \Theta_{\rm NP}^{\, \figeight} (k_\perp/k^-, 1,   \phi_k) = 1$.
The same argument can be repeated for the operator $\overline \Theta_{\rm NP}^{\, \bndry} $ in \eq{thetabndry}. In this case \eq{sdthetapq}  leads to~\cite{Hoang:2019ceu}
\begin{align}
\label{eq:bndry}
\Delta   \overline \Theta_{\rm sd}^{\, cs}  &=\Theta^{\bndry}_k\, \frac{2}{\theta_{cs}}\, 
 \delta \big ( z_{cs} - \tzcut \theta_{cs}^\beta \big)\\ 
&\qquad \times  \frac{1}{Q} \, \big (k^-\, (1 + \beta) - \beta \, k_{\perp} \, \cos( \phi_k) \big) \, ,\nn 
\end{align}
where $\Theta^{\bndry}_k = 1$ when an NP particle is clustered with the subjet with $\overline \Theta_{\rm NP}^{\, \bndry} (k_i^-/k_\perp, 1 , \phi_k) = 1$, and $\Theta^{\bndry}_k = -1$ when the NP particle is lost from the subjet, such that $\overline \Theta_{\rm NP}^{\, \bndry} (k^-/k_\perp, 1 , \phi_k) = 0$.

Note that the shift correction results from an expansion in the $+$ components: $q^+/p_{cs}^+ \ll 1$, whereas the boundary correction from expansion in the $-$ and $\perp$ components: $q_{i}^-/p_i^- \ll 1$ and $q_{i}^\perp/p_i^\perp \ll 1$. In terms of the rescaled momenta $k^\mu$ the perturbative information is factored out in each case as factors of $\theta_{cs}/2$ and $2/\theta_{cs}$ in \eqs{shift}{bndry}, respectively. 

We have shown in \Ref{Hoang:2019ceu} that as a result of these two effects, the leading power corrections to the factorized partonic cross  section in the SDOE region can be cast into the following form
\begin{align}
\label{eq:factfull}
&\frac{d  \sigma_{\kappa}^{\rm had}}{dm_J^2} 
\!= \! \sum_{\kappa={q,g}} \! N_\kappa(\Phi_J,R, \zcut, \beta,\mu)  \\
& \qquad \times \int_0^\infty\!\! d\ell^+    \int_0^\infty \! \! d k   \: J_\kappa \big(m_J^2 - Q\,\ell^+ ,  \mu \big) \nn \\
& \qquad \times \qcut^{\frac{-1}{1+\beta}}\,S_c^\kappa \Big[\big( \ell^+ - C^\kappa_1(m_J^2)k \big) \qcut^{\frac{1}{1+\beta}},\beta,\mu\Big] \nn\\
& \times \bigg(1 - Q\,k \, \frac{d  C^\kappa_1(m_J^2)}{d m_J^2}  + \frac{Q\Upsilon^{\kappa}_1(\beta)}{m_J^2} \, C_2(m_J^2)\bigg)  F^\kappa_{\figeight}(k)  
\, ,\nn
\end{align}
where the shape function $F^\kappa_{\figeight}(k)$ satisfies:
\begin{align}
\label{eq:moments2}
\int_0^\infty d k \>k\,  F^\kappa_{\circ \! \! \circ} (k) =\Omega_{1\kappa}^{\figeight}  \,, \qquad \int_0^\infty d k \> F^\kappa_{\figeight} (k)  =1  \, .
\end{align}
Here $\Omega_{1\kappa}^{\figeight}$ and $\Upsilon^\kappa_1 (\beta)$ are the hadronic parameters related to the shift and boundary power corrections respectively and depend on the partonic channel $\kappa$. The superscript `$\figeight$' for the shift correction is meant to distinguish it from the power correction known from ungroomed event shapes and is a reminder that it results from the specific geometry  of the catchment area shown in \fig{NPfig8}. The  coefficients $C_1(m_J^2)$ and $C_2(m_J^2)$ also depend on additional variables, and have the interpretation of resummed average of the functions of the angles of collinear-soft radiation for a given jet mass $m_J^2$: 
\begin{align}
\label{eq:C1C2sim}
C_1^\kappa(m_J^2)  &= C_1^\kappa(m_J^2, Q, \tzcut, \beta, R) 
 = \frac{\left  \langle \theta_{cs}(m_J^2)/2\right \rangle}
  {\langle 1\rangle(m_J^2)} 
  \, , \\
 C_2^\kappa(m_J^2) &= C_2^\kappa(m_J^2, Q, \tzcut, \beta, R) 
\nn \\
 &= \left\langle\frac{2}{\theta_{cs}(m_J^2)}\, \frac{m_J^2}{Q^2}\,  \delta \big ( z_{cs} - \tzcut \theta_{cs}^\beta \big) \right\rangle 
 \frac{1}{\langle 1\rangle(m_J^2)} 
 \, . \nn
\end{align}
The structure of the factorization formula in \eq{factfull} relies on a LL approximation for deriving the hadronization corrections, which at this order only involve the perturbative coefficients $C_1^\kappa$ and $C_2^\kappa$. The partonic resummation, involving evolution of the scales in the functions $N_\kappa$, $J_\kappa$ and $S_c^\kappa$, is not restricted in this manner. 

We see that a key feature of nonperturbative corrections in \eq{factfull} is the universality property of the hadronic parameters. First, the hadronic parameters $\Omega_{1\kappa}^{\figeight}$ and $\Upsilon^\kappa_1 (\beta)$ are independent of the energy $Q$ and the jet mass $m_J$. The dependence of the power correction on $Q$ and $m_J$ is accounted for by the Wilson coefficients. Further, the parameter $\Omega_{1\kappa}^{\figeight}$ is both $\tzcut$ and $\beta$ independent.  The boundary power correction parameter $\Upsilon^{\kappa}_1(\beta)$ has, however, linear $\beta$ dependence  as seen from \eq{bndry}. Note that from the point of view of the field theory  derivation the function $F^\kappa_{\circ\!\!\circ}$ does not formally contain corrections from underlying event or multiple-parton interactions, but it can serve as a useful model for describing these effects~\cite{Stewart:2014nna}, which we adopt here as well.

In the following sections we show that the same hadronization parameter $\Omega_{1q}^{\figeight}$ appears in the leading non-perturbative corrections for a jet generated by a massive top quark, even when accounting for the top decay products, and that the effect of the boundary corrections is suppressed for top jets in the region of $M_J$ most relevant for top mass measurements.

\section{Hadron Level Factorization for Top Jets}  \label{sec:massive}

In this section we derive a hadron level factorization formula that predicts the top jet mass $M_J$ spectrum for a boosted top quark initiated jet with $Q\gg m_t$ and light soft drop grooming, as illustrated in \fig{expansion}.  We  focus on the peak region of the jet mass spectrum since it is particularly sensitive for measurements of the top mass. The peak region is defined by 
\begin{align}
  M_J^2 - m_t^2 \sim m_t \Gamma \,,
\end{align}
with $\Gamma$ determined by the top width $\Gamma_t$ as well as additional  smearing effects from non-perturbative radiation, i.e. $\Gamma \gtrsim \Gamma_t$. We consider the jet mass range $170\, {\rm GeV} \le M_J \le 190\,{\rm GeV}$  for our analysis. Due to the presence of the top mass the peak of the jet mass spectrum is close to $m_t$, and it is convenient to define the rescaled jet mass variable
\begin{align}
\label{eq:shat}
\hat s_t  \equiv \frac{(M_J^2-m_t^2)}{m_t} 
\,.
\end{align}

\subsection{Partonic Modes for Light Groomed Top Jets} 
\label{sec:topmodes}

In the peak region, the dynamics of the top jet includes the top decay $t \to b W \to b q\bar q'$, where we call the $bq\bar q'$ final state particles the primary top quark decay products. For the dynamics at scales much below the top mass, the top quark and its associated gluon radiation, are described in a strongly boosted version of heavy quark effective theory (HQET) with four velocity $v=(m_t/Q,Q/m_t,0)$, referred to as boosted-HQET (bHQET)~\cite{Fleming:2007qr}. The bHQET modes include in particular the ultra-collinear (UC) modes which are related to the radiation that is soft in the top rest frame.  In this context the dynamics of the top-decay  can be incorporated either differentially by perturbative calculations, or inclusively with the top width $\Gamma_t$. The UC modes have the momentum scaling:
\begin{align}\label{eq:puc}
p_{uc}^\mu \sim \hat s_t\Big(\frac{m_t}{Q}, \frac{Q}{m_t},1 \Big) \, .
\end{align}
Figure~\ref{fig:modesmassive} shows the kinematic location of the UC modes as well as the other modes relevant for using soft drop on a top initiated jet.
The collinear-soft (CS) mode is again located at the intersection of the (blue) measurement line and the (orange) soft drop constraint lines. However its scaling is modified compared to \eq{pCS} due to the presence of $m_t$ by taking $m_J^2\to m_t \hat s_t$, yielding for the $p^\mu = (p^+,p^-,p^\perp)$ components 
\begin{align} \label{eq:pcstop}
  p_{cs}^\mu &\sim
   \hat s_t\, \frac{m_t}{Q\zeta}\, \Bigl( \zeta , \frac{1}{\zeta} , 1\Bigr) 
\,,
\end{align}
where the angular parameter is now
\begin{align} \label{eq:zeta}
  \zeta \equiv \Bigl(\frac{m_t\, \hat s_t}{Q\,\qcut}\Bigr)^{\frac{1}{2+\beta}}
   \,.
\end{align} 
The soft modes (S) appear in a  manner similar to the massless jet case as can be seen by comparing \figs{modesmassless}{modesmassive}.

These perturbative modes describe the top jet mass in the peak region with light grooming through the following parton level factorization theorem:
\begin{align}  \label{eq:factpartonic}
&\frac{d\hat \sigma}{dM_Jd\Phi_J} 
= N(\Phi_J,R, m_t, z_{\rm cut},\beta,\mu)\! 
\\
& \times 
\int\!\! d\ell^+\, 
J_B\Big(\hat s_t- \frac{Q\ell^+}{m_t},\delta m ,\Gamma_t,\mu\Big)   S_c^{q}\Big[  \ell^+ \qcut^{\!\frac{1}{1+\beta}}
,\beta,\mu\Big] \,.
\nn
\end{align}
It resembles \eq{masslessFact0} in the sense that the collinear soft function is the same as for the massless quark initiated jets. The factor $N$ appearing here differs from  \eq{masslessFact0} since it describes $t\bar t$ production, but as before encodes other aspects of the event, including the hard process and the PDF's. We also take $N$ to include the branching fraction for the hadronic top decay $\Gamma_{t\to bq\bar q'}/\Gamma_t$, which accounts for the fact that the jet mass measurement is made on the hadronic decay. It also depends on the top mass $m_t$ scale, but the dominant sensitivity of the cross section to $m_t$ results from the dynamics in the collinear sector. The dynamics of the  UC modes and the inclusive top decay is described by the bHQET jet function $J_B(\hat s_t, \Gamma_t, \delta m, \mu)$. At tree-level $J_B(\hat s_t, \Gamma_t, \delta m, \mu)$ is just a Breit-Wigner function peaked at the top quark mass $m_t$ and thus carries the main mass sensitivity. It also appears in the analysis of ungroomed top jet mass~\cite{Fleming:2007qr,Fleming:2007xt}, is known at two-loop order~\cite{Fleming:2007xt,Jain:2008gb}, and provides control over the renormalization scheme scheme for the top mass. This is indicated by the dependence on
\begin{align}
\label{eq:deltam}
\delta m = m_t^{\rm pole} - m_t \,,
\end{align}
which is understood to be expanded in the perturbation series of $\alpha_s$ for $J_B$.
At leading order in $m_t/Q$, the direction of the top quark is equivalent to that of its decay products. Hence, after integrating out $m_t$, the UC light quark and gluon modes are not affected by the top decay at leading power. This allows us to define the bHQET jet function $J_B(\hat s_t, \Gamma_t, \delta m, \mu)$ for an unstable top in terms of the stable quark jet function $J_{B}(\hat s_t,\delta m,\mu)$ convolved with the top quark Breit-Wigner~\cite{Fleming:2007xt}:
\begin{align}  \label{eq:UnstableJb}
&J_B  (\hat{s}_t, \Gamma_t, \delta m ,\mu )   \\ 
&\qquad =\int_{-\infty}^{\hat{s}_t}\!\!\df \hat{s}^\prime \: 
J_B(\hat{s}_t-\hat s', \delta m, \mu )  
\frac{\Gamma_t}{\pi \big(\hat{s}^{\prime\, 2} + \Gamma_t^2\big)} 
\,. \nn
\end{align} 
In \eqs{puc}{pcstop}, when we consider the threshold limit $\hat s_t\to 0$ the power counting scaling for the UC modes should be considered to saturate at $\hat s_t\to \Gamma_t$ since the top quark width provides an effective infrared cutoff as long as we treat the decay inclusively.

We now discuss the constraints on the soft drop parameters that are related to the expansions used in \eq{factpartonic}. We first note that the dependence on the soft drop parameters $\zcut$ and $\beta$ in \eq{factpartonic} only enters through the collinear-soft function and the normalization $N$ whereas the UC radiation in the bHQET jet function is unaffected by soft drop. This is ensured by demanding that grooming is sufficiently light so as to not affect the UC modes. Hence
\begin{align} \label{eq:sdkeepuc}
\tzcut \theta_{uc}^\beta \ll z_{uc}  &\Rightarrow  \tzcut \Big(\frac{2 m_t}{Q}\Big)^\beta \ll  \frac{\hat{s}_t}{m_t}  \nn \\
&\Rightarrow 
 \zcut \ll \frac{\hat s_t}{m_t} \Big(\frac{p_T}{m_t}\Big)^\beta \, ,
\end{align}
where the subscript `uc' refers to the angles and energy fractions for the UC modes.
We will see below in \sec{topkinematics} that consideration of the kinematics of the top decay products leads to the light grooming constraint stated above in \eq{zcutcons}, which actually is stronger than \eq{sdkeepuc}. An important consequence of the validity of \eq{sdkeepuc} is that it allows us to use the well established description of the UC sector in terms of the bHQET jet function so that we have a full control over the top mass scheme specified by the series $\delta m$ in \eq{factpartonic}.

Next, we demand that the CS modes are sufficiently boosted in the peak region such that they factorize from the global soft modes. This is ensured by taking
\begin{align} \label{eq:csboosted}
  \zeta\ll \frac{1}{\cosh \eta_J} 
     \ \ \Rightarrow\ \
    \zcut^{\frac{1}{2+\beta}}\gg \frac{1}{2} 
   \Big(\frac{\hat s_t}{m_t}\frac{m_t^2}{p_T^2}\Big)^{\frac{1}{2+\beta}}
   \, ,
\end{align}
which with $\hat s_t \gtrsim \Gamma_t$ yields the condition in \eq{zcutcons2}  for the peak region. \eq{csboosted} automatically ensures that the ultrasoft modes (US) with momentum scaling $\hat{s}_t m_t/Q(1,1,1)$, that contribute significantly to the ungroomed jet mass, are groomed away. To see this, we note that $z_{us} = \hat{s}_t m_t/Q^2$ and $\theta_{us} \sim 1$, so \eq{csboosted}  implies $\tzcut > z_{us}$, which is the condition needed to groom away the wide angle ultrasoft modes.

\subsection{Nonperturbative Modes for Light Groomed Top Jets} 

\begin{figure}[t!]
	\raisebox{4.3cm}{a)}
	\includegraphics[width=0.44\textwidth]{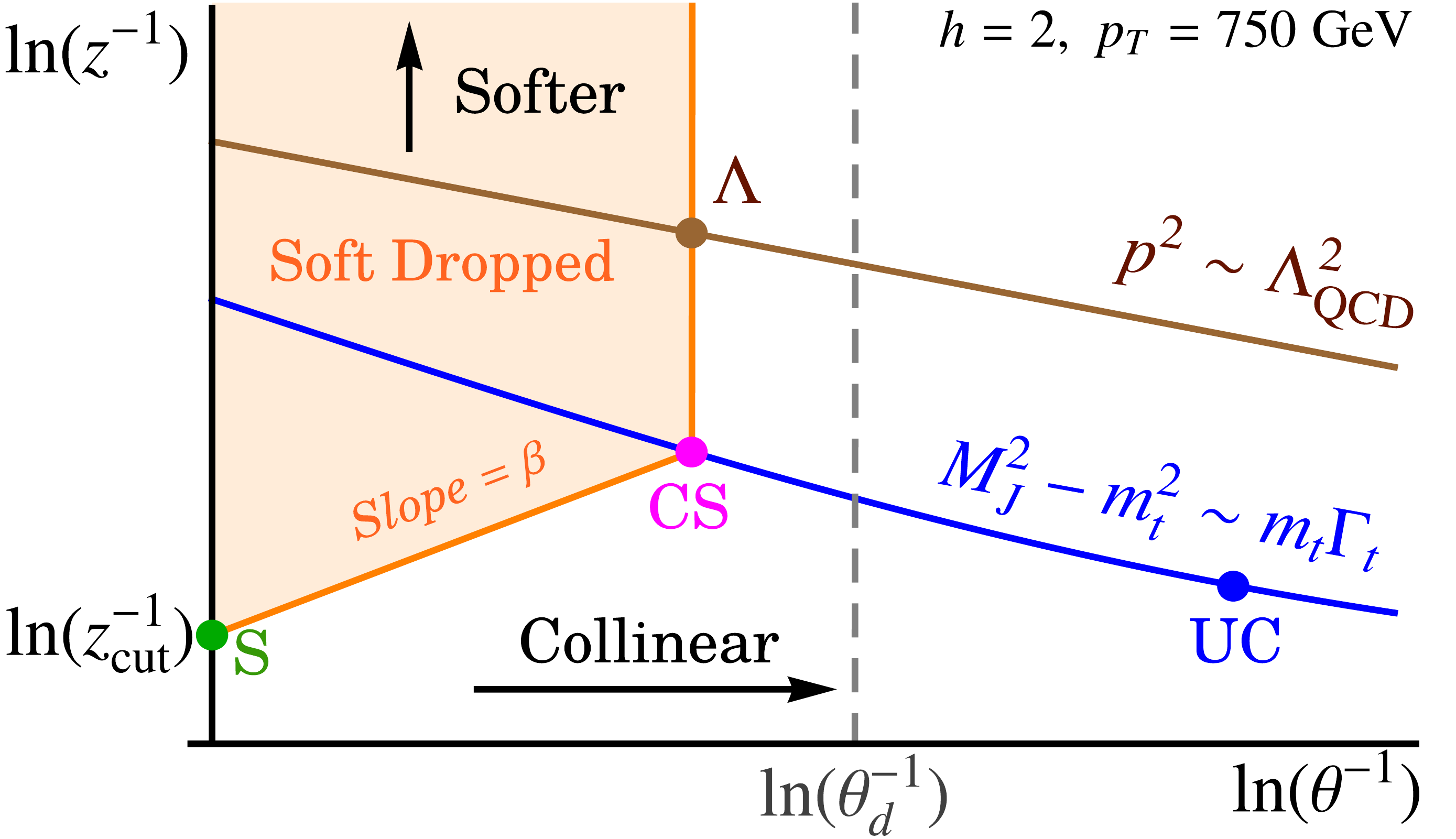}\\[5pt]
	\raisebox{4.3cm}{b)}
	\includegraphics[width=0.44\textwidth]{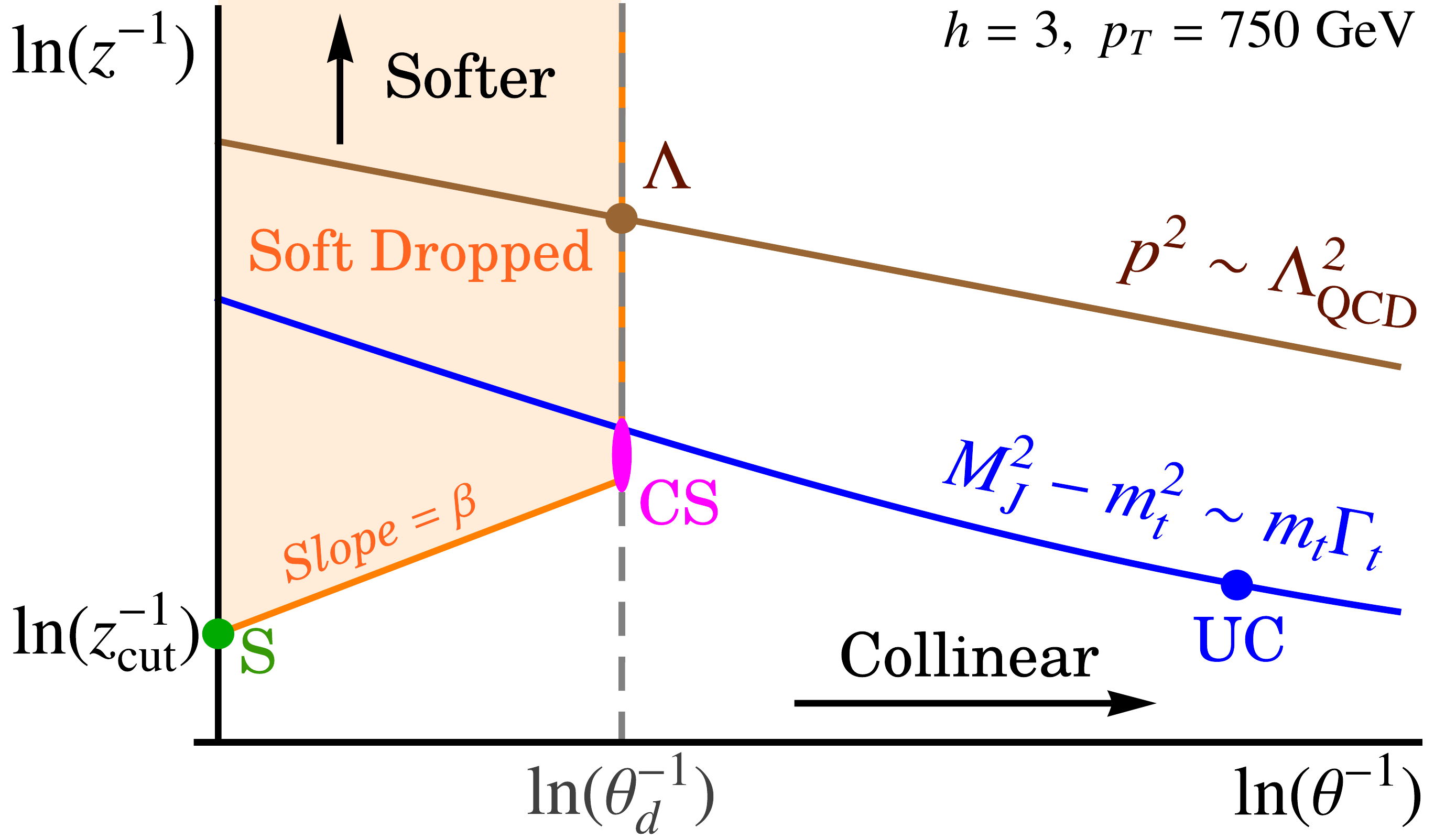}\\
	\vspace{-0.2cm}
	 
	\caption{Relevant SCET modes for soft drop jet mass for a top jet. The dashed vertical line corresponds to the angle of the furthest top-decay product from the jet axis which determines $h$, shown with two different values in (a) and (b).  This leads to a dependence on $h$ in the factorization formula. The other parameters $(p_t, \, \eta_J, \, \zcut, \, \beta)$ are held fixed.
		\label{fig:modesmassive} 
	}
	\vspace{-0.2cm}
\end{figure}

An important aspect when soft drop is applied to a jet containing a highly unstable particle like the top quark, is how the soft drop algorithm stops.  At leading power, with resummation included, there are two possibilities: 
\begin{itemize}
\item[i)]  soft drop stops from a comparison of a collinear-soft subjet and a subjet that contains all the top decay products, 
\item[ii)]  soft drop stops from a comparison between two subjets both of which contain top decay products. 
\end{itemize} 
For massless quark or gluon induced jets the possibility ii) does not arise.
The hadronization corrections arising from the nonperturbative $\Lambda$ modes must be considered separately for these two cases.  It is worth noting that if the progression of the soft drop algorithm through the CA clustering tree reaches a comparison between subjets which each have a decay product then soft drop will always stop due to the large energy fraction carried by the decay products at leading power in the light grooming regime. In \fig{modesmassive}a we show a case where i) is realized, which we will refer to below as the `high-$p_T$'' contribution to the factorization. In \fig{modesmassive}b we show a case where ii) is realized, which we will refer to as the ``decay'' contribution to the  factorization.   The key new ingredient necessary to decide whether case i) or case ii) arises is the decay  angle $\theta_d$ which we  now define.

In the CA clustering tree formed as part of the soft drop algorithm, the subjets (or particles) are ordered and grouped by their pairwise angular distances $R_{ij}$. Because we are in the small angle limit $R_{ij}\ll 1$, we have $R_{ij} \simeq \theta_{ij} \cosh(\eta_J)$, so it is fine to make all comparisons with the relative angles $\theta_{ij}$ where $\cosh(\eta_J)$ scales out as an overall factor. The decay angle $\theta_d$ is determined at a stage in the CA tree where two of the three top decay products from $t\to bq\bar q'$ are grouped into a single subjet, and compared with a second subjet that contains the third decay product.  $\theta_d$ is then defined as the polar angle relative to the top-jet axis of the subjet from this pair that points furthest away from the top-jet axis. To be concrete, it is given by
\begin{align} \label{eq:thetad}
	\theta_d \equiv \max \Bigl(\tilde \theta_{(xy)t}, \tilde \theta_{zt} \Bigr) \,.
\end{align}
Here we use `$xy$' to refer to the pair $q\bar q'$, $qb$, or $\bar q'b$ that is closest in angle, which are grouped together in a subjet, such that
\begin{align}
\tilde \theta_{xy} = \min\Bigl( \tilde\theta_{q\bar q'}, \tilde\theta_{qb}, \tilde\theta_{\bar q'b} \Bigr)  \,,
\end{align}
with $\tilde\theta_{q\bar q'}, \tilde\theta_{qb}, \tilde\theta_{\bar q'b}$ being the pairwise angles between the three decay products. Then the angle $\tilde \theta_{(xy)t}$ in \eq{thetad} is measured between the top decay axis and the parent $xy$-subjet obtained by adding the  momenta $p_x^\mu+p_y^\mu$ following the standard jet recombination scheme. The third particle we call $z$ lies in the other subjet and $\tilde\theta_{zt}$ in \eq{thetad} is its angle relative to the top decay axis. 
For case ii) soft drop stops due to the comparison of these two subjets. We can also consider the impact of the polar angle of the closer subjet containing decay products given by 
\begin{align} \label{eq:thetadp}
\theta_d' &\equiv \min \bigl(\tilde \theta_{(xy)t}, \tilde \theta_{zt} \bigr) \, .
\end{align}
However, as we show below in \sec{universality}, its effects can be neglected in comparison to $\theta_d$ for the kinematic region of interest.

Since the two decay subjets just discussed are dominated by the kinematics of the energetic decay products, $\theta_d$ can be calculated considering the $t\to b q \bar q'$ tree level decay and its phase space variables $\Phi_d$:
\begin{align}  \label{eq:thetadphid}
  \theta_d=\theta_d(\Phi_d,m_t/Q) \,.
\end{align} 
Here we define $\Phi_d$ as the 5 independent dimensionless kinematic variables  of the top-decay $t\to bq\bar q'$ in its rest frame (4 angles and one energy fraction). The dependence on $m_t/Q$ arises from boosting to the frame where the top quark has high $p_T$ and where the angles entering the soft drop conditions are computed. The function $h$ which appeared in \eq{zcutcons0}, and which enters the decay component of the factorization theorem, is 
directly related to $\theta_d$ through the definition
\begin{align}
\label{eq:hdef}
\tan\Bigl( \frac{\theta_d}{2}\Bigr) = \frac{m_t}{Q}\, h\Bigl(\Phi_d, \frac{m_t}{Q}\Bigr) \,. 
\end{align}
The $m_t/Q$ prefactor on the RHS pulls out the dominant dependence that $\theta_d$ has on the boost of the top quark, indicating that $\theta_d\to 0$ as $Q/m_t\to \infty$. 
For the decay component illustrated in \fig{modesmassive}b the collinear-soft modes still have the scaling in \eq{pcstop}, but here
\begin{align} \label{eq:pcsdtop}
   \zeta \sim \zeta_d \equiv \frac{m_t}{Q}\, h\Bigr(\Phi_d,\frac{m_t}{Q}\Bigr)  \,.
\end{align}
The use of the tangent in \eq{hdef} makes the $p^+$ component of modes at this angle scale in a manner proportional to $h$ without further approximations, since $p^+ \propto  \tan(\theta_d/2)$.

Hence, the scaling of the leading nonperturbative modes in the two scenarios is given by
\begin{align} \label{eq:pLtop}
& \text{i)\phantom{i} high-$p_T$:} 
& &p_\Lambda^\mu \sim \Lambda_{\rm QCD} \Bigl( \zeta , \frac{1}{\zeta}, 1\Bigr) \nn
\, ,\\
& \text{ii) decay:} 
& &p_\Lambda^\mu \sim \Lambda_{\rm QCD} \Bigl( \zeta_d , \frac{1}{\zeta_d}, 1\Bigr) 
\, .
\end{align}
Examples for the geometry of the modes are shown in \fig{modesmassive}a for case i) and in Fig.~\ref{fig:modesmassive}b for case ii),  by considering two different values for $h$, implying two different values of $\theta_d$. Figure~\ref{fig:modesmassive} implies that the factorization theorems for a jet initiated by the unstable top quark has a  significantly different structure compared to the massless quark jet case. From \eq{pLtop} we see that hadronization corrections from the $\Lambda$ modes depend on tests of kinematic information of the perturbative cross section components. Whether we are in the case i) or ii) will depend on the relative sizes of $\theta_{cs}$ 
and the decay angle $\theta_d$. In particular,  for the case $\theta_{cs}\sim \theta_d$ the top decay may affect  the definition of the collinear-soft function, 
denoted as $S_c^{(d)}$, in a non-trivial way, causing it to differ from \eq{masslessFact0}.
However, we will show below in \sec{topkinematics} that this difference can be neglected in the light grooming region at NLL order.

In general, resolving the comparison between the angles $\theta_{cs}$ and $\theta_d$ requires kinematic information about the CS modes, as well as kinematic information about  the decay products considering the phase space $\Phi_d$ and the boost $Q/m_t$. In the next section we show how appropriate expansions for the factorization formula can be obtained by generalizing both the collinear-soft function $S_C$ and the bHQET jet function $J_B$ for ultra-collinear radiation to account for the dependence on the decay product phase space, while still remaining inclusive over the primary decay products in the jet mass measurement.

\subsection{Incorporating Top-Decay Product Kinematics} \label{sec:topkinematics}

To obtain results accounting for both cases i) and ii) when soft drop acts on a boosted top quark jet, we consider the collinear-soft function $S_c^{(d)}$ that in addition to the soft drop condition also accounts for the angular information of the decay products.  To be concrete, for the calculation of $S_c^{(d)}$ one now has to account for the vertical dashed line in \fig{modesmassive}b at $\theta = \theta_d$ which represents an additional phase space boundary of the soft drop region, and hence affects the location of the CS mode. At one loop order, working in the region $\zeta\sim \zeta_d$ as indicated by \fig{modesmassive}b, one finds that the resulting collinear-soft function $S_c^{(d)}$ has the form
\begin{align}  \label{eq:SCd}
& S_c^{(d)}(\ell^+,Q_{\rm cut},\beta,\theta_d,\mu) 
\ \ = S_c^q\Big(\ell^+ Q_{\rm cut}^{\frac{1}{1+\beta}},\beta,\mu\Big)
   \nn \\
& -\frac{\alpha_s(\mu) C_F}{(2\!+\!\beta)\pi} 
\frac{2^{3+\beta}}{\qcut\theta_d^{2+\beta}}\:
{\cal L}_1\biggl(  \! \frac{\ell^+}{\qcut} \frac{2^{2+\beta}}{\theta_d^{2+\beta}} \!\biggr) 
\Theta\biggl[  \frac{\qcut\theta_d^{2+\beta}}{2^{2+\beta}} - \ell^+\! \biggr]
\nn\\
& + {\cal O}(\alpha_s^2) 
\,,
\end{align}
where ${\cal L}_1(x)=[(\ln x)/x]_+$ is the standard logarithmic plus function, see for example Ref.~\cite{Ligeti:2008ac}. Since at one loop in fixed order perturbation theory we have a single emission with momentum $k$, the relevance of high-$p_T$ or decay scenario boils down to a simple comparison between the emission angle $\theta_k$ and $\theta_d$. The phase space integral can be rewritten to have two terms, one with the same form as for massless jets yielding $S_c^q$, and one $\theta_d$-dependent term yielding the term in the second line of \eq{SCd}. The latter term is $\mu$ independent, and involves the combination ${\cal L}_1(x) \Theta(1-x)$ where its $\Theta$ function makes 
\begin{align}
 \Bigr( \frac{\theta_{cs}}{2} \Bigl)^{2+\beta} 
  \simeq  \frac{\ell^+}{Q_{\rm cut}}  
  <\: \Bigr( \frac{\theta_d}{2}\Bigl)^{2+\beta} \simeq 
  \Bigl[ \frac{m_t}{Q} h\Bigl(\Phi_d, \frac{m_t}{Q}\Bigr) \Bigr]^{2+\beta}
 \,.
\end{align}
This ensures that it only contributes in the decay case ii) of \eq{pLtop}. This term involves a logarithm, which is not large when $x\sim 1$.

The term on the second line of \eq{SCd} involves a logarithm that is, however, not large when $\theta_d\sim \theta_{cs}$, which is the region where an explicit comparison between the cases i) and ii) is necessary. Demanding that the argument of ${\cal L}_1(x)$ satisfies $x\gtrsim 1$ ensures that this term either vanishes or is not a large logarithm, and leads to the constraint in \eq{zcutcons} when we set $\ell^+$ to $\hat s_t m_t/Q$ following the scaling determined by \eq{factpartonic}. We saw above in \eq{sdkeepuc}, that this is also necessary to ensure that the UC modes are not affected by soft drop. 
Thus,  when case ii) applies, the term in the second line of \eq{SCd} only enters beyond NLL order for the light grooming region.   Hence, at NLL, the perturative CS function is the same as the one that appeared for massless quark initiated jets in \eq{masslessFact0},
\begin{align}
S_c^{(d)}(\ell^+,Q_{\rm cut},\beta,\theta_d,\mu) \Big|_{\rm NLL}  
= S_c^q\Bigl(\ell^+ Q_{\rm cut}^{\frac{1}{1+\beta}},\beta,\mu\Bigr)
\Big|_{\rm NLL}
.
\end{align}
The decay angle $\theta_d$ is thus only relevant for determining nonperturbative corrections, as we discuss further below.
\begin{figure}[t!]
	\centering
	\includegraphics[width=.98\columnwidth]{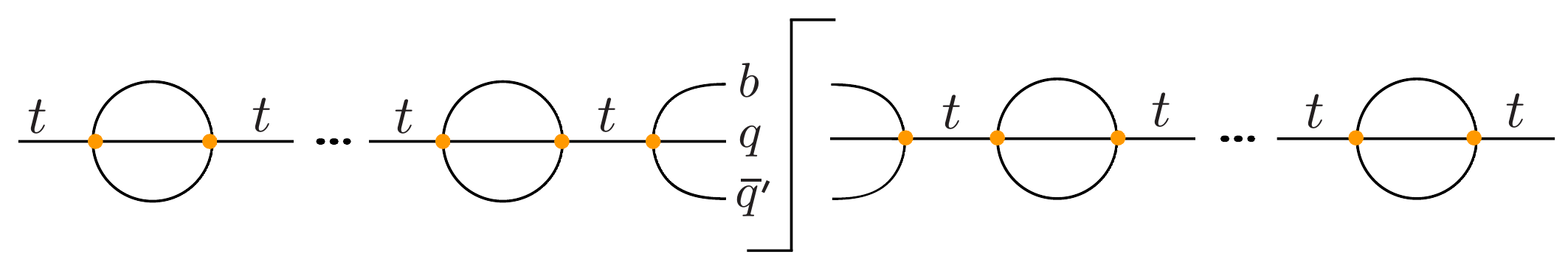} 
	 
	\caption{Bubble chain for an unstable top quark leading to a Breit-Wigner together with a differential distribution for the top decay products. From the closed two-loop bubble calculation involving $b q\bar q'$ only the imaginary top width term is kept.
		\label{fig:topbubblechain}}
\end{figure}

Next we discuss how to account for the $\Phi_d$ phase space dependence of the decay angle $\theta_d$.  Since this dependence is determined at the scale $m_t$, where the top decay takes place, it can be thought of as differential hard matching information from the perspective of the low energy CS, UC, and S modes.  Hence, in the light grooming region the description of the dynamics of the UC modes, which cannot resolve details concerning the top decay, is still based on the inclusive stable heavy quark jet function $J_B(\hat s, \delta m, \mu)$ in \eq{UnstableJb}.

We refer to the function that encodes the distribution of the top decay products as $D_t(\hat s',\Phi_d,m_t/Q)$ where $\hat s'=(p_t^2-m_t^2)/m_t$ is the offshellness of the top-quark defined in analogy to \eq{shat}. The function $D_t(\hat s',\Phi_d,m_t/Q)$ and the stable jet function $J_B(\hat s, \delta m, \mu)$ together account for all leading order effects concerning the top decay and the UC modes. The calculation of $D_t$ involves a geometric sum of top self energy bubbles, where one hadronically decaying bubble is cut, as shown in \fig{topbubblechain}. In the non-cut bubbles we just keep the total imaginary top width $\Gamma_t$. We can then write the result in terms of the purely resonant contribution
\begin{align} \label{eq:Dt}
D_t\Big(\hat s', \Phi_d,\frac{m_t}{Q}\Big)  
&=  \frac{ \Gamma_t }{\pi(\hat s^{\prime\,2} \!+\! \Gamma_t^2)} \,d_t\Bigl(\Phi_d,\frac{m_t}{Q}\Bigr)
\Bigl[ 1 \!+\!  {\cal O}\Big( \frac{\hat s'}{m_t}\Big) \Bigr]
 .
\end{align} 
The full  3-body phase space dependence of the hadronic top-decay $t\to b W\to b q\bar q'$ is contained in the dimensionless decay function
\begin{align} \label{eq:dt}
d_t\Big(\Phi_d,\frac{m_t}{Q}\Big) = 
\frac{1}{\Gamma_{t\to b q\bar q'}^{\cal J}} \frac{d\Gamma_{t\to bq\bar q'}}{d\Phi_d}  \,,
\end{align}
where $d_t$ also includes the full kinematics of the possibly resonant $W$ propagator,  and is normalized such that 
\begin{align} \label{eq:dtnorm}
\int_{\cal J}\! d\Phi_d\: d_t\Bigl(\Phi_d,\frac{m_t}{Q}\Bigr) = 
 1
\,.
\end{align}
The hadronic width $\Gamma_{t\to b q\bar q'}^{\cal J}$ in \eq{dt} and the normalization condition in \eq{dtnorm} are both defined with a phase space cut indicated by the superscript or subscript ${\cal J}$. The precise definition of this phase space restriction is described below.
Recall that there is a factor of the branching ratio $\Gamma_{t\to bq\bar q'}/\Gamma_t$ in the cross section normalization factor $N$ in \eq{factpartonic}.
The residual dependence on the boost factor $m_t/Q$ in $d_t$ is still important and hence indicated explicitly in \eq{dt}, while we leave implicit its dependence on $m_W/m_t$.
The expansion in \eq{Dt} indicates that for the description of the angular dependence of the decay products we can work with an onshell quark decay function with $\hat s'=0$. 

The formula for $D_t$ in \eq{Dt} indicates that the dependence on the top quark's off-shellness $\hat s'$ appears in a Breit-Wigner, which factorizes from the $\Phi_d$ dependent decay function $d_t$ that is generated at the scale $m_t \gg \hat s'$.  As a result, on convolving the stable jet function with $D_t$ one recovers the unstable jet function in \eq{UnstableJb} upon integrating over $\Phi_d$.

\begin{figure}[t!]
	\includegraphics[width=0.45\textwidth]{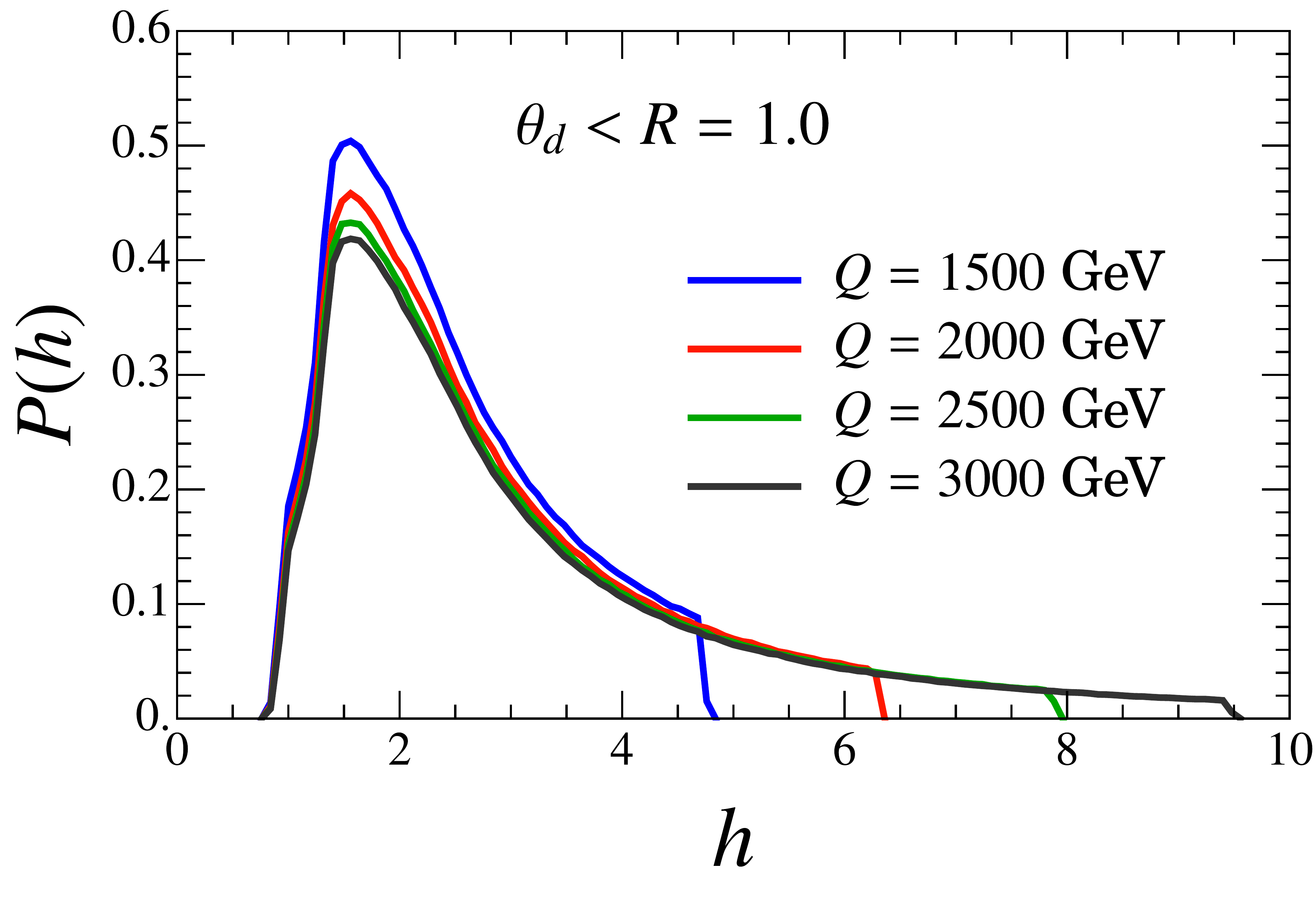}
	\vspace{-0.2cm}
	 
	\caption{Probability distribution of $h$ for different values of $Q$ with $m_t=173.1\,{\rm GeV}$.
		\label{fig:hhist} 
	}
	\vspace{-0.2cm}
\end{figure}

The key information contained in the decay function $d_t$ needed for the hadron level factorization formula of \eq{hprimedef} is encoded in the distribution of the values of the function $h(\Phi_d,m_t/Q)$, which is the basis of implementing the information on the decay angle $\theta_d$ defined in \eq{thetad} in \eq{hdef}. This distribution has the form
\begin{align}	\label{eq:Phdef}
&P\Big(\tilde h,\frac{m_t}{Q}\Big) 
 =\frac{1}{\Gamma_{t\to b q\bar q'}^{\cal J}} \frac{d\Gamma_{t\to bq\bar q'}}{d \tilde h} \nn \\
 &\qquad = \int_{{\cal J}} \! d\Phi_d \, 
 d_t\big(\Phi_d,\frac{m_t}{Q}\big) \: 
 \delta\Big(\tilde h \!-\! h(\Phi_d, \frac{m_t}{Q})  \Big) 
 \, ,
\end{align}
and is shown in \fig{hhist} for different $Q$ values. It peaks at $h$ values near $2$. As anticipated it drops to zero near $h=1$. For larger $h$ the distribution falls off, and then drops sharply to zero. The drop off occurs because we demand that the decay products are always contained within the original (ungroomed) jet of radius $R$, which gives an upper limit on $\theta_d$.  This is indicated by the subscript ${\cal J}$ in \eq{Phdef}. 
While this restriction for boosted tops is technically related to a power correction to the constraint of \eq{dtnorm}, we find it useful to include it explicitly in \eq{Phdef} since it ensures that only decay products that are actually inside the original ungroomed jet can stop soft drop. Since this $R$ dependence is quite mild we have suppressed it in the arguments of $P(\tilde h,m_t/Q)$. We account for this phase space restriction in all $\Phi_d$ integrals, including the normalization imposed by \eq{dtnorm}, which ensures that
\begin{align}
  \int d\tilde h\  P\bigl(\tilde h, m_t/Q \bigr) = 1 \,.
\end{align}
 
\subsection{Universality with Light Quark Jets}
\label{sec:universality}

Having determined the distribution $P(\tilde h,m_t/Q)$, and thus through \eq{thetadphid} the distribution of the decay angle $\theta_d$, we can now address how the relevant scenarios of \eq{pLtop} are implemented in the hadron level factorization formula.
Using the angle $\theta_d'$ defined in \eq{thetadp}, we define a corresponding function $h'$ via
\begin{align} \label{eq:hprimedef}
  \tan\Bigl( \frac{\theta_d'}{2}\Bigr) &= \frac{m_t}{Q}h'\Bigl(\Phi_d, \frac{m_t}{Q}\Bigr) \,. 
\end{align}
In general we have $h > 1$ and $h' < 1$, and most often $h'\ll h$.  Since $m_t/Q \ll 1$, at the stage of the clustering where the comparisons in \eq{thetad} are made, the $xy$ and $z$ subjets containing top decay products carry almost all of the jet-momentum. Hence at leading order in the power counting they are in the same plane as the total jet 3-momentum vector, and the angle between the $xy$ and $z$ subjets is $\theta_d+\theta_d'$.  This angle sum has to be compared with the angles $\theta_{cs}$ of the  CS subjets encountered in traversing the clustering tree backwards that could have large enough $p_{T}$ and $R_{ij}$ to stop the soft drop grooming via \eq{SD}. In analogy to \eq{Phdef} we can define the probability distribution of the function $h'$, $P(h')$.  Using $P(h')$ we then find from an explicit computation that $\langle h'\rangle/\langle h\rangle \simeq 0.22$ for $Q/m_t=5$. This ratio decreases to $\langle h'\rangle/\langle h\rangle \lesssim 0.12$ when $Q/m_t\ge 11$, which is the range of interest for our analysis. Therefore for our treatment of the $\Lambda$ modes we can safely assume $\theta_d' \ll \theta_d$ or equivalently $h'\ll h$. This implies that the determination of which of the cases in \eq{pLtop} applies can be found by a simple comparison of which subjets are at a wider angle relative to the top jet axis:
\begin{align} \label{eq:thetaCSorD}
& \text{case i)\phantom{i} if } \theta_{cs} > \theta_d  
\,, 
\\
& \text{case ii) if } \theta_{cs} < \theta_d    
\,.
\nn
\end{align}
Here we limit ourselves to LL resummation for the determination of $\theta_{cs}$.
\begin{figure}[t!]
	\includegraphics[width=0.3\textwidth]{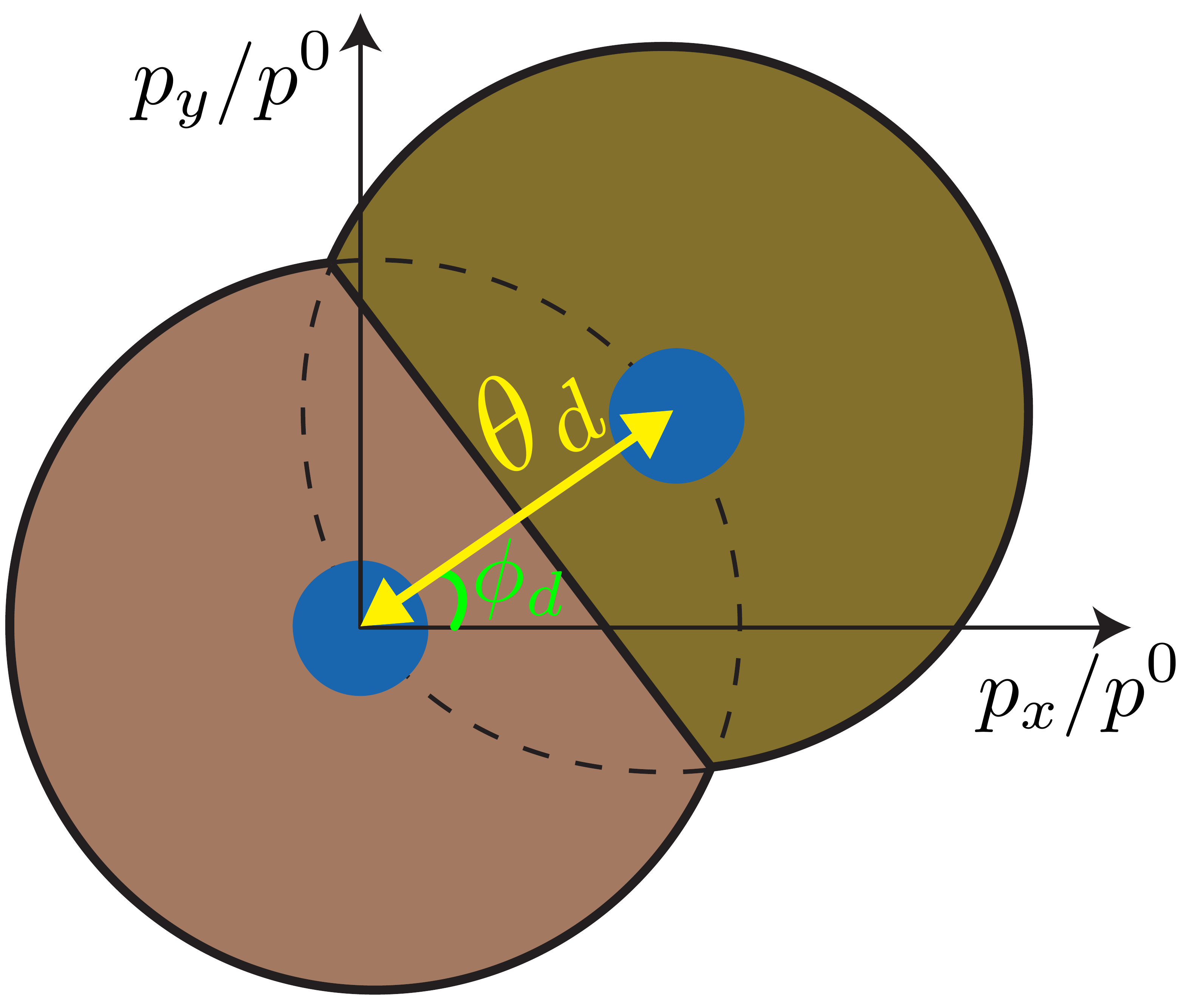}\\
	\vspace{-0.2cm}
	 
	\caption{The non-perturbative modes kept by the soft drop factorization theorem at NLL when the groomer is stopped on comparison between top decay products (blue dots), pictured from above looking down the jet axis. 
		\label{fig:NPfig8_Decay} 
	}
	\vspace{-0.2cm}
\end{figure} 

Following the discussion \sec{masslessfact}, with $\theta_d^\prime \ll \theta_d$  the geometry of the catchment area of the nonperturbative modes for the shift term when the top decay products stop soft drop is illustrated 
in \fig{NPfig8_Decay}. Thus we see that the catchment areas in the high-$p_T$ and the decay factorization cases have the same geometry.
In contrast to \fig{NPfig8} the blue dots here signify that both the subjets containing top decay products are much more energetic than the typical collinear-soft subjets formed by pure QCD radiation. An important consequence is that the boundary correction originating from modifications of the soft drop test of the energetic $xy$ and $z$ subjets due to nonperturbative modes is strongly suppressed by $\Lambda_{\rm QCD}/m_t$.  We will come back to discussion of the boundary corrections for the high-$p_T$ case at the end  of \sec{tophad}.

The projection operator for the shift term given in \eq{thetashift} for massless jets now simply generalizes to the following expression:
\begin{align}
\label{eq:thetashiftmax}
\overline \Theta_{\rm NP}^{\, \figeight}(  p_{\Lambda }^\mu, \, &\theta_{X}, \phi_{X})  = \Theta\bigg(| \Delta \phi| -  \frac{\pi}{3} \bigg)\Theta\bigg(1 -\frac{ \theta_{\Lambda}}{\theta_{X}} \bigg) \nn\\
&+ \Theta\bigg(\frac{\pi}{3} - | \Delta \phi| \bigg)\Theta \bigg(2 \cos (\Delta \phi) - \frac{\theta_{\Lambda}}{\theta_{X}}\bigg)\,  
\end{align}
where $\theta_X = \max\{\theta_d, \theta_{cs}\}$. The nonperturbative factorization for boosted tops thus involves the rescaling of the nonperturbative momenta in analogy to \eq{rescaleNP} via the angle $\theta_X$ and we find that the shift correction is parametrized by $\Omega_{1q}^{\figeight}$ given in \eq{moments2}, which is the shift correction parameter for massless quark jets.

Thus, the main conclusions of this analysis for the decay case are (a) that for the shift correction the catchment area, as shown in \fig{NPfig8_Decay}, has the same geometry as for the high-$p_T$ case, described already for the massless quark initiated jets in  \fig{NPfig8}, and (b) that the boundary corrections can be neglected.

\subsection{Comparing High-$\mathbf{p_T}$ and Decay Components}\label{sec:tophad}
\begin{figure}[t!]
	\includegraphics[width=0.45\textwidth]{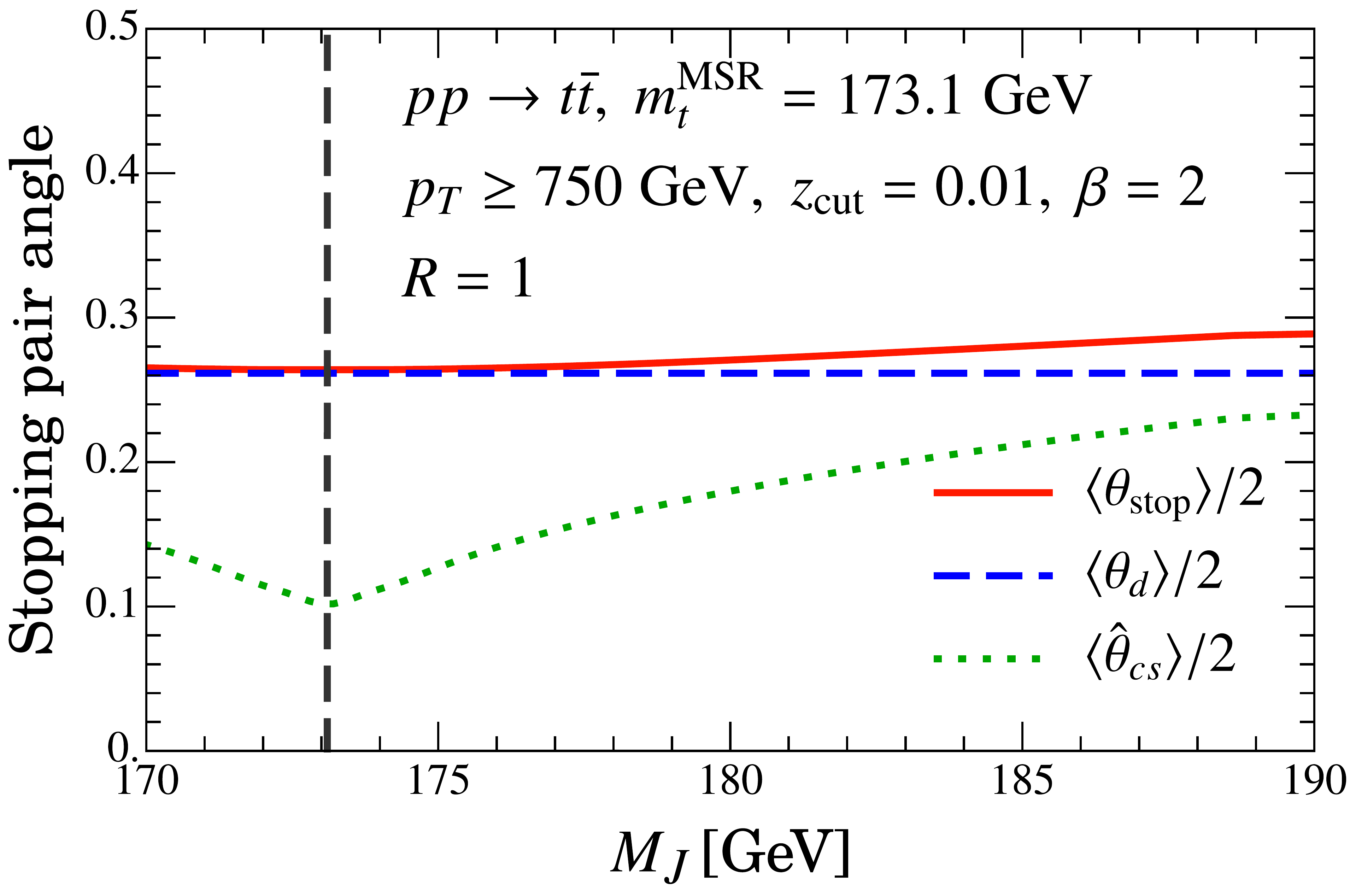}
	\vspace{-0.2cm}
	 
	\caption{Average $\langle\theta_d\rangle$, $\langle \hat\theta_{cs} \rangle$ and the winning stopping pair angle, $\langle\theta_{\rm stop}\rangle$, as a function of top jet mass. 
		\label{fig:thetad} 
	}
	\vspace{-0.2cm}
\end{figure} 

We can use the results derived above to dynamically determine for $\theta_d'\ll \theta_d$ the relative fraction of the events where soft drop stops either on a subjet containing a top decay product, or on a gluon radiation initiated collinear-soft subjet. 
 
 From \eq{C1C2sim} we note that the average (resummed) opening angle $\theta_{cs}$ is simply twice the coefficient $C_1$ with $m_J^2$ replaced by $m_t\hat s_t$ in the peak region:
\begin{align}
\label{eq:thetaCS}
 \langle \hat \theta_{cs} \rangle (\hat s_t)
 &\equiv  2\, C_1^{q(pp)}(m_t|\hat s_t|) 
 \\
 & \equiv 2 \,C_1^{q(pp)} \big(m_J^2 = |m_t \hat s_t|,  Q, \tilde z_{\rm cut}, \beta, R\big) 
 \, . \nn
\end{align} 
We give the result for $C_1^{q(pp)}(m_J^2, Q, \tilde z_{\rm cut}, \beta, R)$ in \App{NPcoeff}~\cite{Hoang:2019ceu}. An approximate formula for $C_1^{q(pp)}$ that is accurate at the level of 5\% and sufficient for our numerical analysis below is:
\begin{align}
\label{eq:C1approx}
C_1^{q(pp)} (m_J^2) &\simeq \frac{0.55}{2 \cosh\eta_J} \Big(\frac{\zcut}{0.1}\Big)^{0.12(1 -0.25 \beta)} \\
&\qquad \times\biggl( \frac{m_J^2}{p_T^2 \zcut R_0^{-\beta}} \biggr)^{\frac{1}{2.2+0.8 \beta}} \, . \nn
\end{align}
To determine whether the decay or high-$p_T$ scenario applies, one can simply compare the angle for decay $2m_t\tilde h/Q$ to $2 C_1^{q(pp)}$. 

To get an idea about which scenario dominates we can consider the average opening angle of the stopping pair at a given top jet mass, $M_J$, given by
\begin{align}
\label{eq:thetawin}
 \langle \theta_{\rm stop} \rangle (\hat s_t) &= \,\int d \tilde h\: P\Big(\tilde h, \frac{m_t}{Q} \Big)  \\
&\qquad \times \max\Bigl \{ 
  \frac{2m_t}{Q}\, \tilde h , \, 
  \langle \hat \theta_{cs}\rangle (\hat s_t) \Bigr\} \, , \nn
\end{align} 
and compare it to $\langle \hat \theta_{cs}\rangle (\hat s_t)$ and the average decay angle
\begin{align}
\langle \theta_d \rangle =  \, \int d \tilde h\: 
 P\Big(\tilde h, \frac{m_t}{Q} \Big)\:
 \frac{2m_t}{Q} \, \tilde h  \, .
\end{align}
Note that unlike $\langle \hat \theta_{cs}\rangle$ in \eq{thetaCS}, $\langle \theta_d \rangle$ is computed entirely  from perturbative dynamics at the top mass scale $m_t$, and can be obtained from a fixed-order calculation.
\fig{thetad} shows the averages  $\langle\theta_d\rangle $ (dashed blue line) and $\langle \hat \theta_{cs}\rangle$ (dotted green), and the average winning angle $\langle \theta_{\rm stop}\rangle$ (solid red line) as a function of the jet mass for $\zcut = 0.01$, $\beta = 2$, and $p_T \geq 750$ GeV. The dashed vertical line is at $M_J = m_t = 173.1$ GeV which is the input top quark mass.  This kinematic point and choice of grooming parameters satisfies the light grooming constraints in \eq{zcutcons0}. 
\begin{figure}[t!]
	\includegraphics[width=0.45\textwidth]{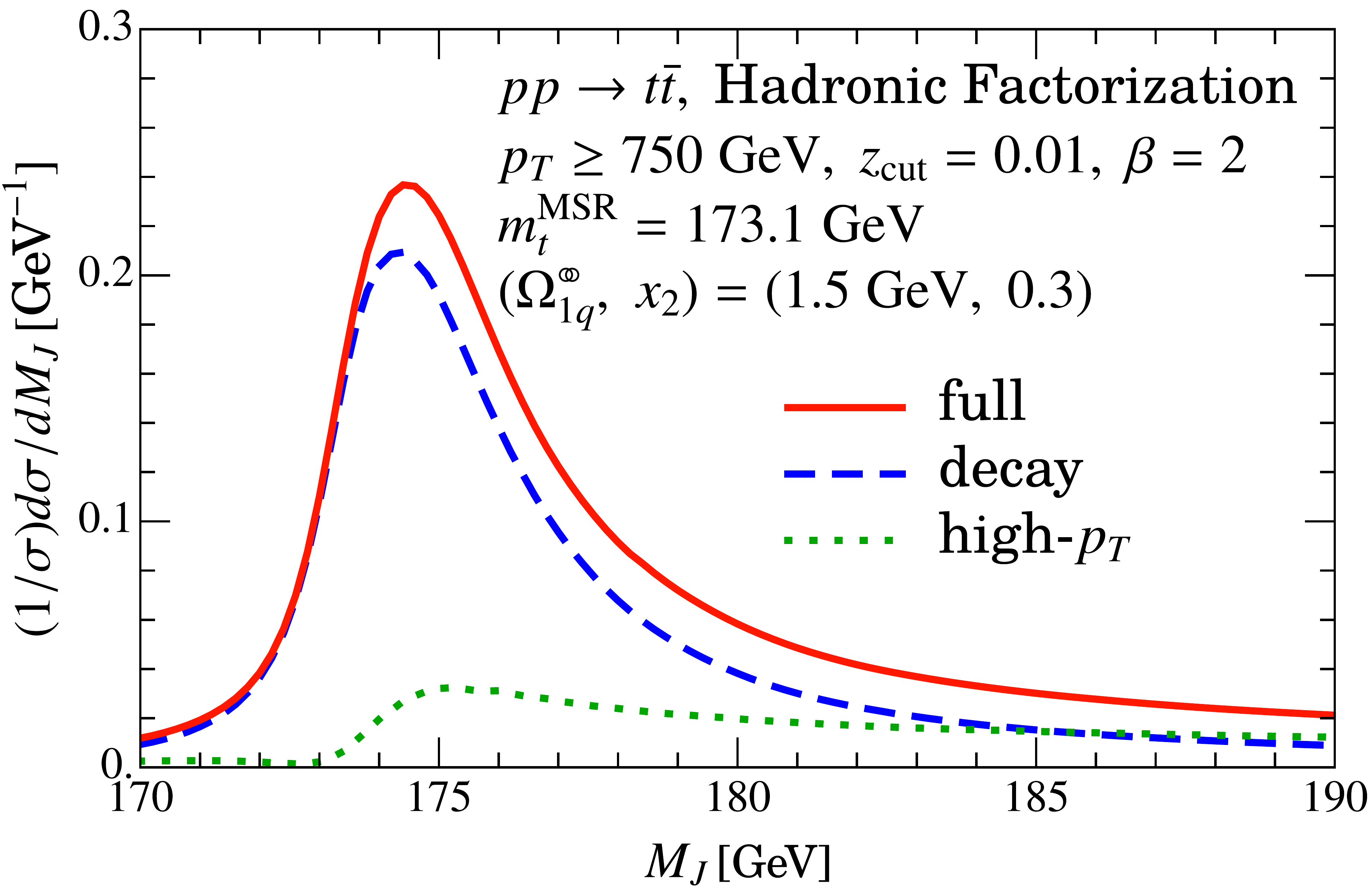}
	\vspace{-0.2cm}
	 
	\caption{The full NLL hadronic top jet mass cross section along with the decay and high-$p_T$ components for representative values of the top mass $m_t^{\rm MSR}$ in the MSR scheme and the hadronic parameters.
		\label{fig:thcomp} 
	}
	\vspace{-0.2cm}
\end{figure} 
We see that in the peak region soft drop is predominantly stopped by the top decay products. Cases where the groomer is stopped by a collinear-soft subjet, which corresponds to the the high-$p_T$ scenario, become significant only at higher jet masses in the tail of the spectrum above the peak region.

It is even more conclusive to carry out the same type of analysis at the cross section level. Even though we present the hadron level factorization theorem later in \eq{factthmfinalNLL}, we briefly analyze here its decay and high $p_T$ components, shown in \fig{thcomp}.;
We use a representative MSR top mass $m_t^{\rm MSR}(R_m = 1 \,{\rm GeV}) = 173.1$ GeV as input, and the hadronic parameters $(\Omega_{1q}^{\figeight}, x_2)$ = (1.5 GeV, 0.3), where $x_2$ is related to the second moment of the nonperturbative shape function (defined below in \eq{x2def}). 
We again see that the decay component dominates, but that there is a significant contribution from the high-$p_T$ component in the tail of the spectrum.

\begin{figure}[t!]
	\includegraphics[width=0.45\textwidth]{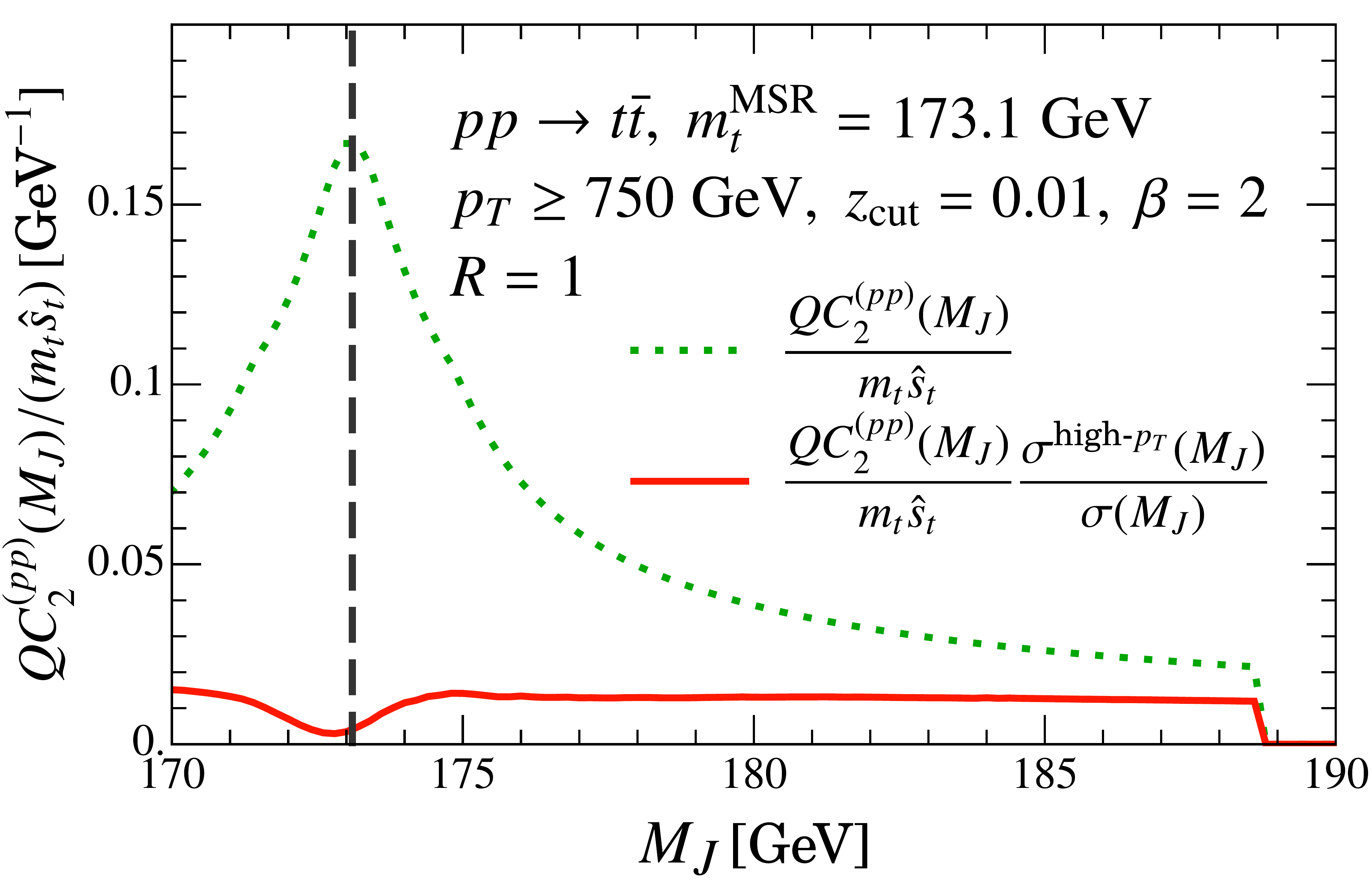}
	\vspace{-0.2cm}
	 
	\caption{The nonperturbative coefficient $C_2^{q(pp)}Q/(m_t\hat s_t)$ that appears for boundary correction. The actual correction to the jet mass cross section is significantly smaller on account of reduced contribution from high-$p_T$ case.
		\label{fig:C2} 
	}
	\vspace{-0.2cm}
\end{figure} 

As we have pointed out already in \sec{universality} for the decay scenario only the shift contribution to the hadronization corrections has to be accounted for, while in the high-$p_T$ scenario, both boundary and shift corrections are present. However, for decaying top quarks, the boundary terms are highly suppressed as we now show.
Following \eq{factfull}, the boundary hadronization correction to the high-$p_T$ component of the cross section is given by multiplying the parton level cross section by $C_2(M_J) \Upsilon_1^q(\beta)  Q/(m_t \hat s_t)$.   In \fig{C2} we show $C_2(M_J)Q/(m_t \hat s_t)$ as a function of the jet mass $M_J$. For $| \Upsilon_1^q (\beta) | \sim 1$ GeV, the relative correction to the high-$p_T$ is $\sim$ 15\% in the peak region. However, taking into account the reduced overall contribution of the high-$p_T$ component in the peak region, as shown in \fig{thcomp}, the overall correction amounts to less than 2\% for the entire jet mass spectrum.  Therefore we conclude that the only relevant effect from hadronization to the top jet mass is the shift correction and the boundary correction can in general be ignored. 
Interestingly, this implies that all the leading nonperturbative corrections to the top jet mass are described by the same parameter $\Omega_{1q}^{\figeight}$. Note that the higher order power corrections for the shift contributions, $\Omega_{nq}^{\figeight}$, are more important than boundary corrections, since $[p_T \Lambda_{\rm QCD}/(m_t\hat s_t)]^n$ is larger than 2\%. They can be modeled by the first few higher order moments of $F_{\figeight}^q(k)$, 
\begin{align}
\label{eq:momentsn}
  \Omega_{nq}^{\figeight}  \equiv \int_0^\infty \!\! dk \: k^n \: F_{\figeight}^q(k)
  \,.
\end{align}

\subsection{Top Factorization Formula with Soft Drop} \label{sec:topfact}

Building on the results obtained in the previous sections we are now at the stage to extend the partonic factorization formula in \eq{factpartonic} to hadron level.  The final expression for the hadron level factorization theorem for boosted top quark jet mass distribution with soft drop is
\begin{align}  \label{eq:factthmfinalNLL}
&\frac{d\sigma^{\rm NLL}(\Phi_J)}{dM_J} 
= N(\Phi_J,z_{\rm cut},\beta,\mu)\! 
\int \!\! d\tilde h \ P\Big(\tilde h,\frac{m_t}{Q}\Big) 
\\
&\ \ \times 
\int\!\! d\ell^+\, 
J_B\Big(\hat s_t- \frac{Q\ell^+}{m_t},\delta m ,\Gamma_t,\mu\Big)   
\int\!\! dk^+ \:  F^q_{\circ\!\!\circ}(k^+)  
\nn \\
&\ \ \times S_c^{q}\Bigg[ \!\bigg( \ell^+\!\!\! 
-  \!\max\biggl\{\! C_1^{q(pp)}(m_t \hat s_t),
\frac{m_t\tilde h }{Q}  \biggr\} k^+\! \bigg) \qcut^{\!\frac{1}{1+\beta}}
,\beta,\mu\Bigg] \nn \\
&\ \  \times \bigg\{1 - \Theta\Bigl(C_1^{q(pp)}(m_t \hat s_t) - \frac{m_t \tilde h}{Q}\Bigr)\frac{Qk^+ }{m_t}\frac{d C_1^{q(pp)}(m_t \hat s_t)}{d \hat s_t}\bigg\}
 .\nn
\end{align}
This is the top quark analogue of \eq{factfull} that applied for massless  quark and gluon initiated jets. 
The nonperturbative corrections are incorporated by comparing $\theta_{d}/2$ and $\theta_{cs}/2$ at each point in the top decay phase space as shown by the max-function appearing in the argument of the collinear-soft function $S_c^q$.
We note that the normalization correction in the last line due to the derivative of $C_1^{q(pp)}$ appears only for the high-$p_T$ case, i.e.\ when $\theta_{cs}>\theta_{d}$. We also remind the reader that the factorization formula in \eq{factthmfinalNLL} is valid when the light grooming constraints in \eqs{zcutcons}{zcutcons2} are satisfied, and that the derivation of the $C_1^{q(pp)}$ coefficient relied on a LL approximation for formulating the leading power correction. For the partonic cross section there is in principal no restriction on the perturbative order to which the factorization theorem can be applied, but for our analysis we will restrict ourselves to NLL order.

\eq{factthmfinalNLL} involves the same shape function $F^q_{\circ\!\!\circ}(k^+)$ that also appeared for massless quark initiated jets in \eqs{factfull}{moments2}. This emphasizes the fact that the same nonperturbative parameter $\Omega_{1q}^{\figeight}$ appears in both cases, through the moment constraint in \eq{moments2}. 
So in this sense the dominant hadronization effects are universal.
 Thus, at this order the leading nonperturbative power correction hadronization parameter is independent of $p_T$, $\eta_J$, $z_{\rm cut}$, $\beta$, $m_t$, as well as other kinematic variables like those of the top decay.  The dependence on these parameters can be obtained solely using perturbation theory. Furthermore, we note that when considering the nonperturbative corrections for the decay contribution in \eq{factthmfinalNLL} the $Q/m_t$ boost factor in the argument of $J_B$ cancels against the $m_t/Q$ factor multiplying $k^+$ in $S_c^q$. This implies that for the lightly groomed top initiated jet there is a reduced $Q$ dependence of the peak position compared with the ungroomed case. This is also supported by the Monte Carlo studies carried out below and means that the observable peak of the groomed jet mass distribution is substantially closer to $m_t$ than for the ungroomed case.

Of course, by charge conjugation symmetry \eq{factthmfinalNLL} also applies for anti-top initiated jets. In practice, $M_J$ associated to either the hadronically decaying $t$ or $\bar t$ can be measured, while the other can decay hadronically or semi-leptonically. In fully hadronic decays both jets can in principle be sampled independently.  We have explicitly confirmed that our factorization formula in \eq{factthmfinalNLL} satisfies renormalization group consistency for the anomalous dimensions of the various functions at NLL order. Since the resummation analysis is standard we do not discuss it in detail here, but we note that the relevant renormalization scales appearing in the functions in \eq{factthmfinalNLL} after including RG evolution are the obvious combinations of those used in the analyses in Refs.~\cite{Fleming:2007xt,Frye:2016aiz}.

As mentioned in \sec{topmodes}, the dominant top quark mass dependence and the control over the renormalization scheme of the top quark mass sensitivity in \eq{factthmfinalNLL} is contained in the bHQET jet function $J_B$. At NLL order $J_B$ is incorporated at tree-level so that we do not account for any mass scheme corrections $\delta m$ as defined in \eq{deltam}. At this level, the top quark pole mass scheme is implemented if one treats $m_t$ as a fixed and renormalization scale independent parameter. The scale-dependent MSR short-distance top mass scheme $m_t^{\rm MSR}(R_m)$~\cite{Hoang:2008yj,Hoang:2017suc} is implemented by setting $m_t=m_t^{\rm MSR}(R_m)$, and
at NLL order one has to account for the leading logarithmic $R$-evolution of the MSR mass~\cite{,Hoang:2017suc} from a reference scale for which we adopt $R_{m,0}=1$~GeV. Consistency requires that one sets $R_m$ equal to the renormalization scale used in the bHQET jet function $J_B$.
So when we quote an input MSR mass entering the factorization theorem of \eq{factthmfinalNLL} we quote the reference value $m_t^{\rm MSR}(R_{m,0}=1\,{\rm GeV})$.

\section{Results} \label{sec:results}

In this section we analyze the results for groomed top jet mass spectra, both with \Pythia simulations and with our hadron level factorization formula in \eq{factthmfinalNLL} valid at NLL order.
\begin{figure}[t!]
\raisebox{4.8cm}{a)}	
\includegraphics[width=0.44\textwidth]{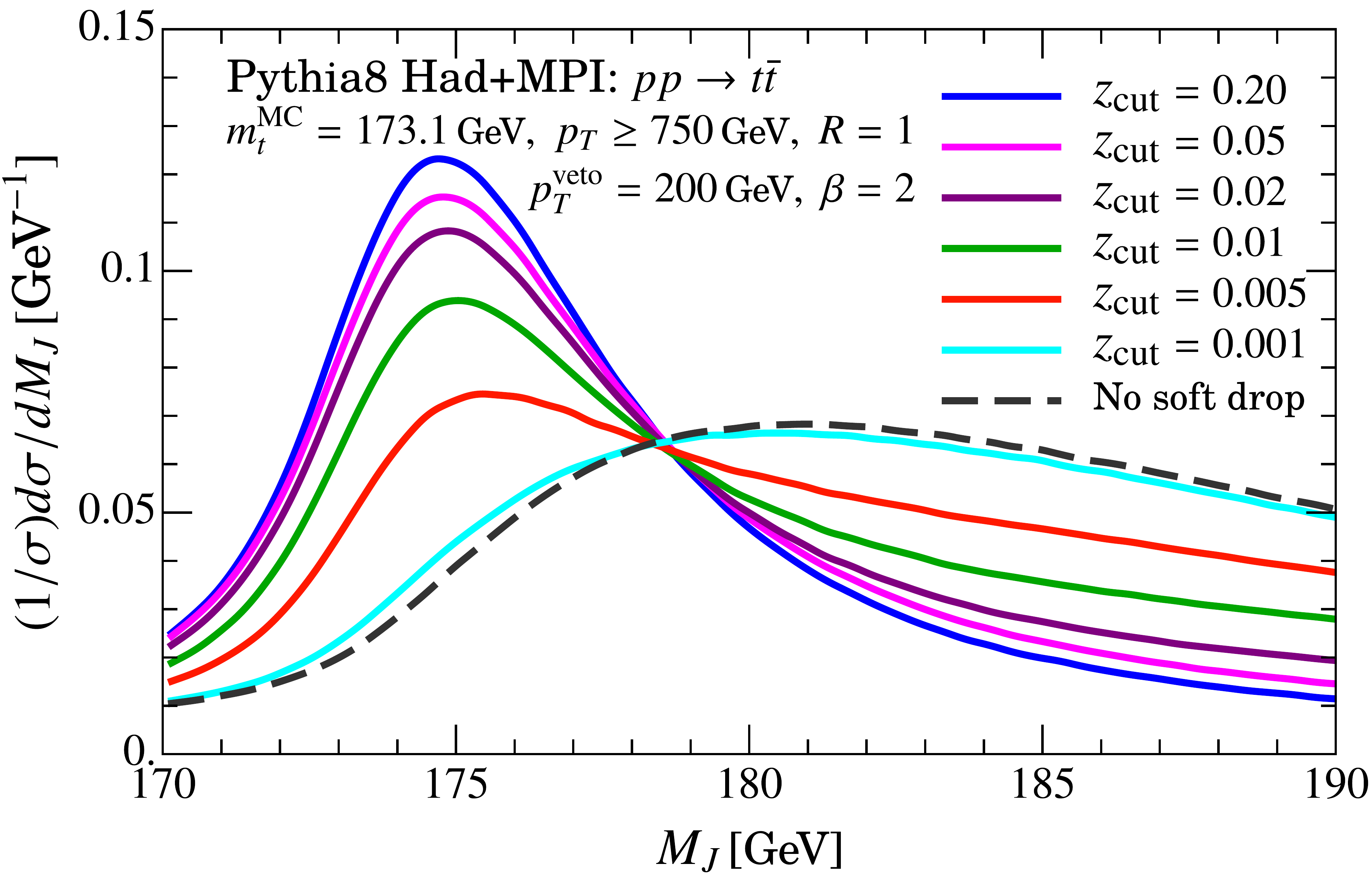}
\raisebox{4.8cm}{b)}
\includegraphics[width=0.44\textwidth]{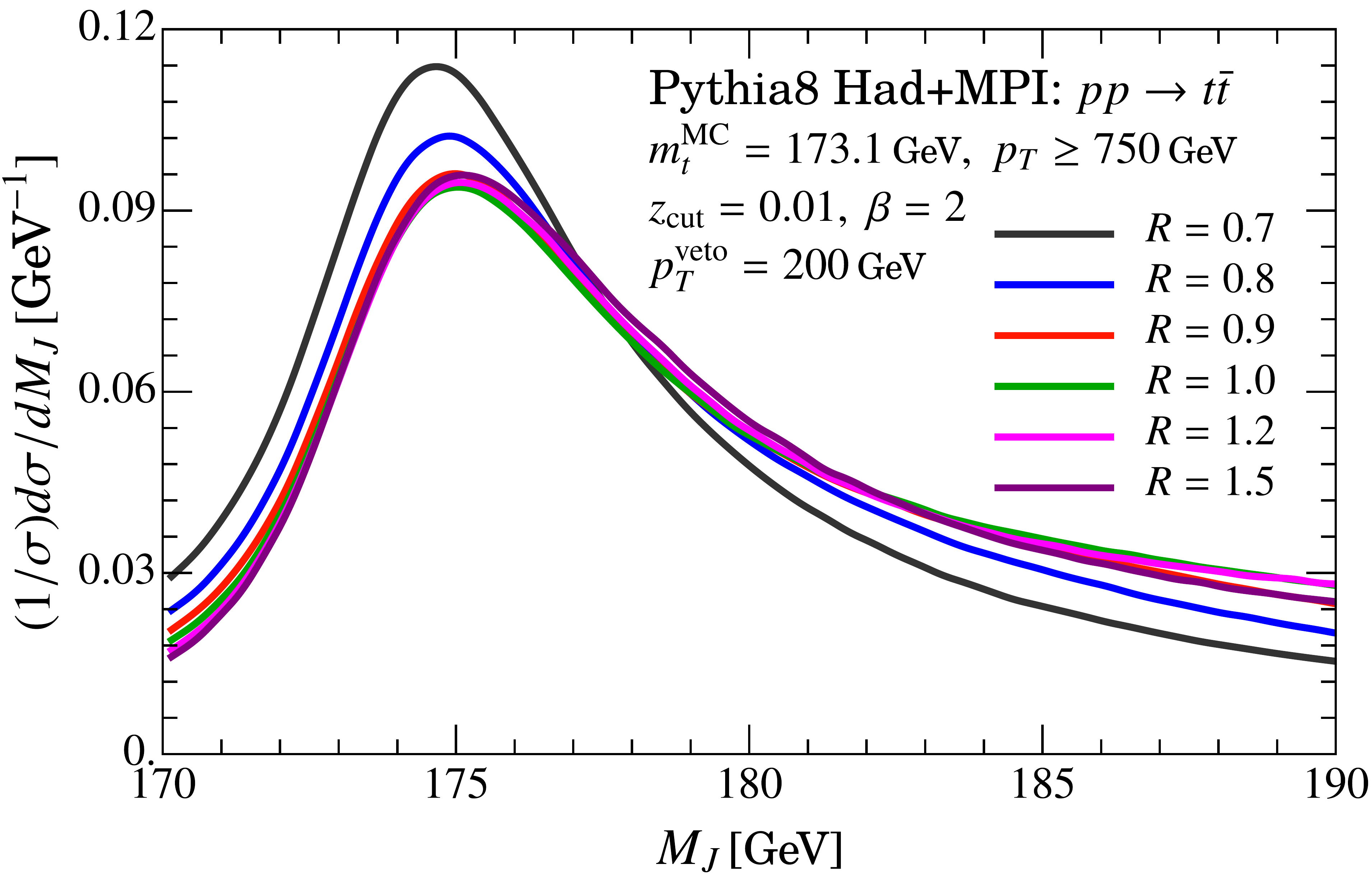}
\raisebox{4.8cm}{c)}
\includegraphics[width=0.44\textwidth]{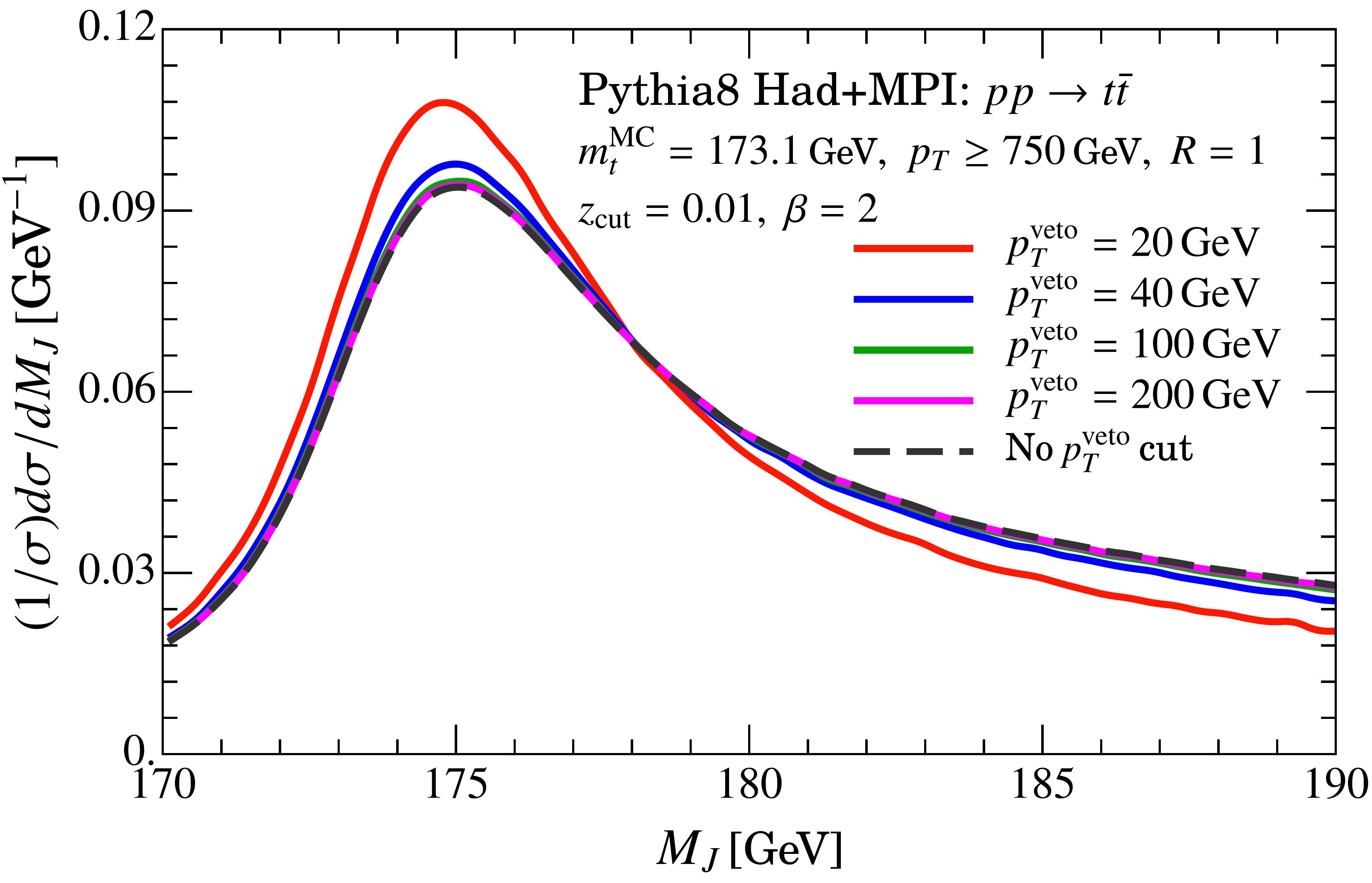}
\raisebox{4.8cm}{d)}
\includegraphics[width=0.44\textwidth]{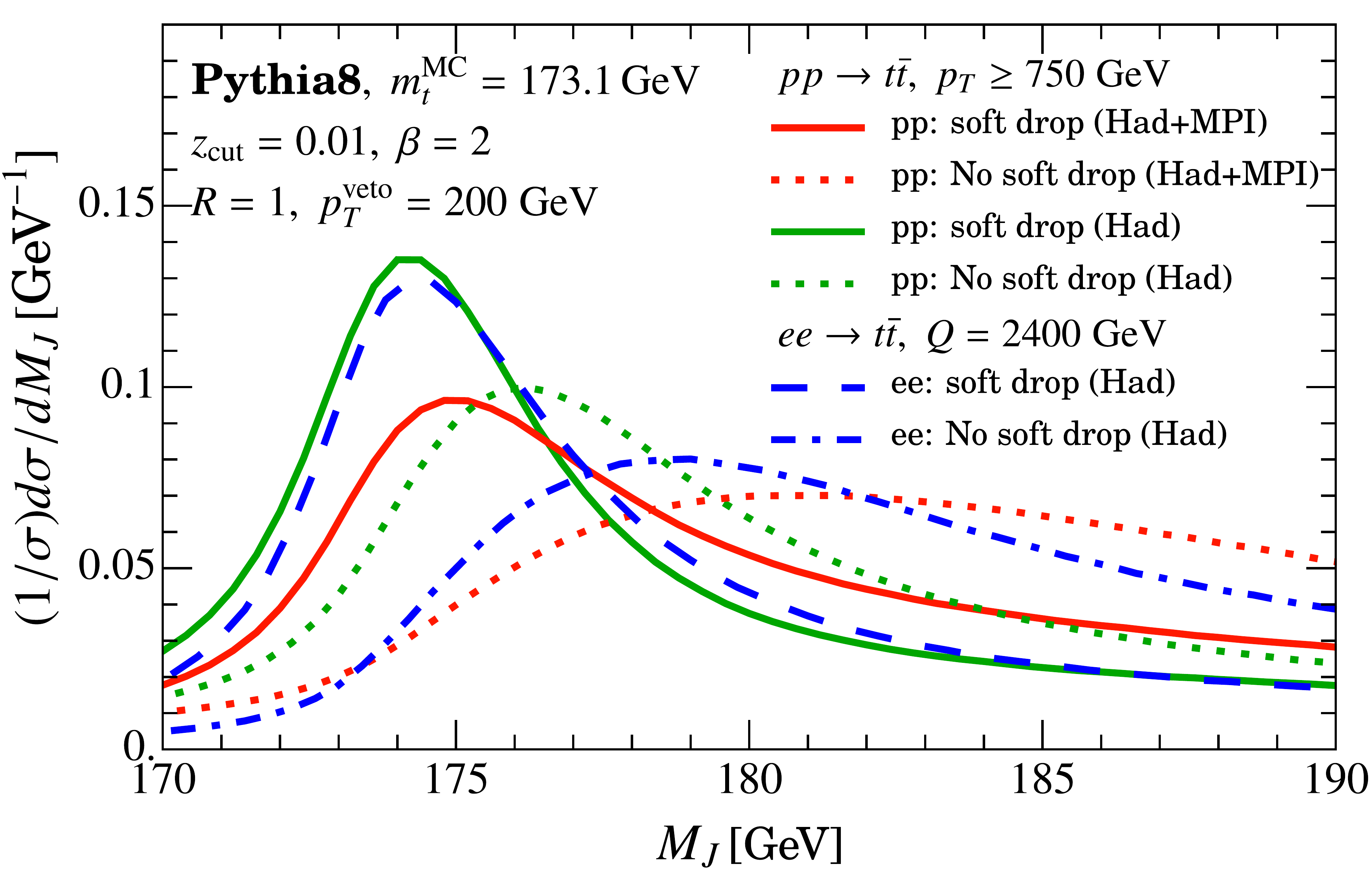}
\vspace{-0.2cm}
\caption{Dependence in \Pythia8 of the $M_J$ spectrum on the a) soft drop parameter $z_{\rm cut}$, b) jet radius $R$, and c) anti-$k_T$ jet veto $p_T^{\rm veto}$. In d) we compare results for $e^+e^-\to t\bar t$ and $pp\to t\bar t$ with and without the light soft drop and with MPI interactions on and off. 
\label{fig:pythiaresults} }
\vspace{-0.4cm}
\end{figure}
At this order, all fixed order matrix elements are evaluated at tree-level and the factorization formula determines the $M_J$ spectrum as a Breit-Wigner distribution dressed by perturbative corrections from resummed large Sudakov double logarithms arising from the hierarchy $p_T \gg m_t \gg \Gamma_t > \Lambda_{\rm QCD}$ and smeared by non-perturbative corrections. As a default for our analysis we take  $p_T\ge 750\,{\rm GeV}$, $|\eta_J|<2.5$, $z_{\rm cut}=0.01$, $\beta=2$, $R_0=1$, based on jets with radius $R=1$ and $p_T^{\rm veto}=200\,{\rm GeV}$.

In Fig.~\ref{fig:pythiaresults} we show groomed jet mass distributions obtained from \Pythia 8.235
including hadronization and MPI effects, and the soft drop plugin in \Fastjet~\cite{Cacciari:2011ma,Larkoski:2014wba}, in order 
to test a number of key features predicted by the factorization theorem.
All curves are normalized to unity over the displayed range.

In Fig.~\ref{fig:pythiaresults}a the dependence on $z_{\rm cut}$ is displayed. We observe that for $z_{\rm cut}\gtrsim 0.005$ the light grooming is effective, as predicted by the constraint in \eq{zcutcons}. Increasing $z_{\rm cut}$ further does not groom soft radiation inside the radius determined by the top decay products, leaving the peak position quite stable even beyond the limit in \eq{zcutcons}, unlike the strong $z_{\rm cut}$ dependence observed for for massless jets~\cite{Larkoski:2014wba}. 

In Fig.~\ref{fig:pythiaresults}b we demonstrate that the light groomed spectrum becomes independent of the jet radius $R$ for $R\gtrsim 0.9$, as expected, in contrast to the strong dependence on $R$ that is present for ungroomed jets~\cite{Dasgupta:2012hg,Jouttenus:2013hs}.  The light groomed spectrum is also independent of an anti-$k_T$ jet-veto cut $p_T^{\rm veto}$ (for jets beyond the two with largest $p_T$) once $p_T^{\rm veto}\gtrsim 50\,{\rm GeV}$, as shown in Fig.~\ref{fig:pythiaresults}c. The independence of the normalized groomed jet mass spectrum with respect to $R$ and $p_T^{\rm veto}$ are predicted by our factorization formula which is independent of these two parameters. The independence to $R$ occurs since we fixed $R_0=1$, and would not be true if one picked $R_0=R$.

Another important prediction of the factorization formula for light groomed top initiated jets is an insensitivity to the parts of the event outside the groomed top jet.  Thus the same factorization theorems apply for top-jets from $e^+e^-\to t\bar t$ and $pp\to t\bar t$, with appropriate modifications related to the definition of $Q$ and the norm function $N$. To obtain a reasonable comparison we take the $e^+e^-$ center-of-mass energy $Q = 2400\,{\rm GeV}$ to approximate the spectrum weighted average $Q$ for $pp$ with  $|\eta_J|<2.5$ and $p_T\ge 750\,{\rm GeV}$. In the $e^+e^-$ case we apply soft drop to top jet hemisphere masses.  The curves displayed in Fig.~\ref{fig:pythiaresults}d show that the spectra for $pp$ and $e^+e^-$ differ substantially without soft drop (dotted green and dot-dashed blue curves, respectively), but agree quite well with soft drop (solid green and dashed blue curves, respectively). Also shown is the impact of MPI on the $pp$ spectra. Without soft drop, adding MPI shifts the peak of the spectrum by $4.5\,{\rm GeV}$ (dotted red versus dotted green), whereas with light soft drop the shift amounts to only $1.1\,{\rm GeV}$ (solid red versus solid green). Formally, effects from UE are outside our framework of factorization, but we can still account for them by extending our treatment for hadronization, which we clarify below.

\begin{figure}[t!]
	\raisebox{4.4cm}{a)}
	\includegraphics[width=0.44\textwidth]{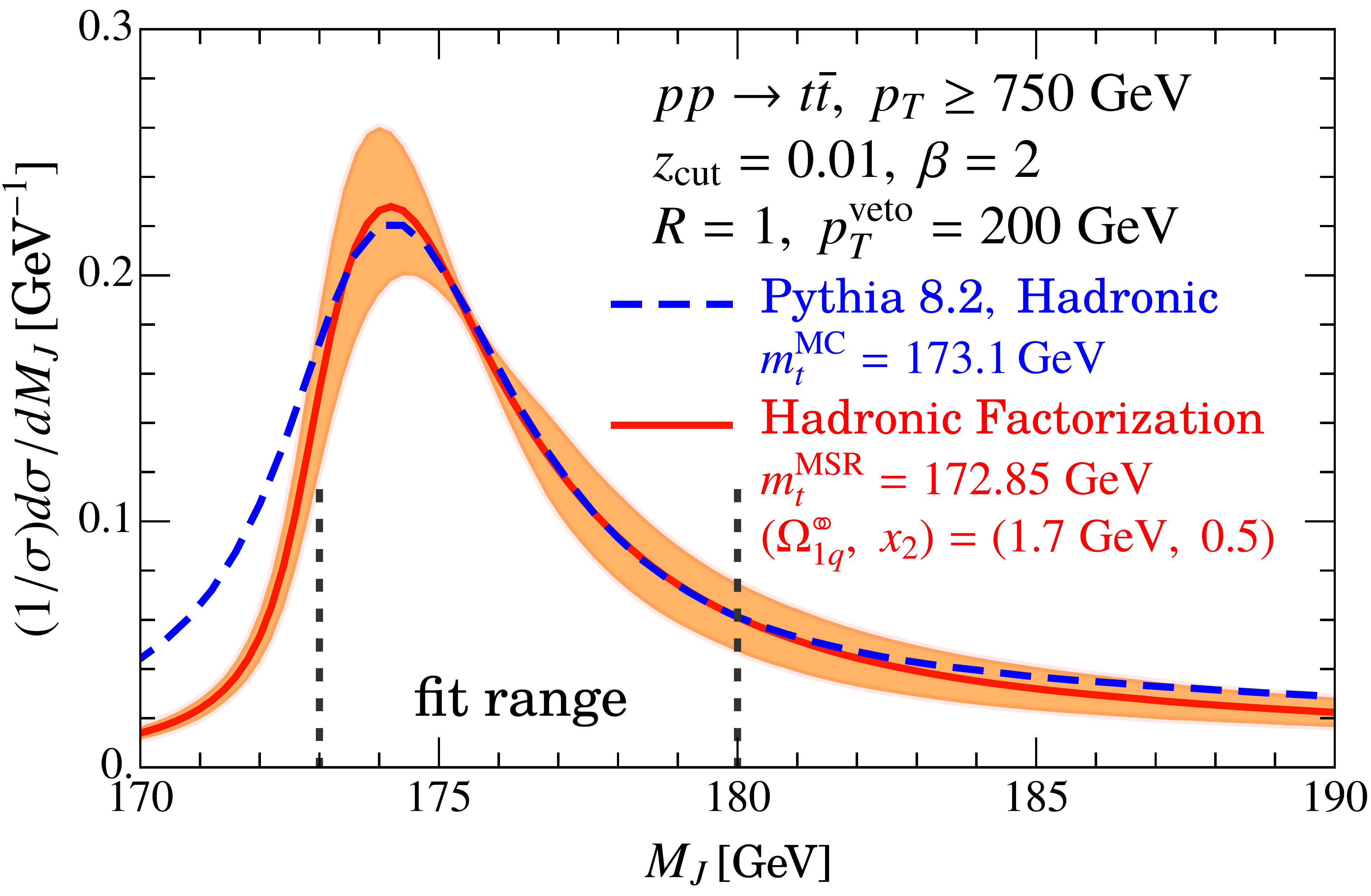}
	 
	\raisebox{4.4cm}{b)}
	\includegraphics[width=0.44\textwidth]{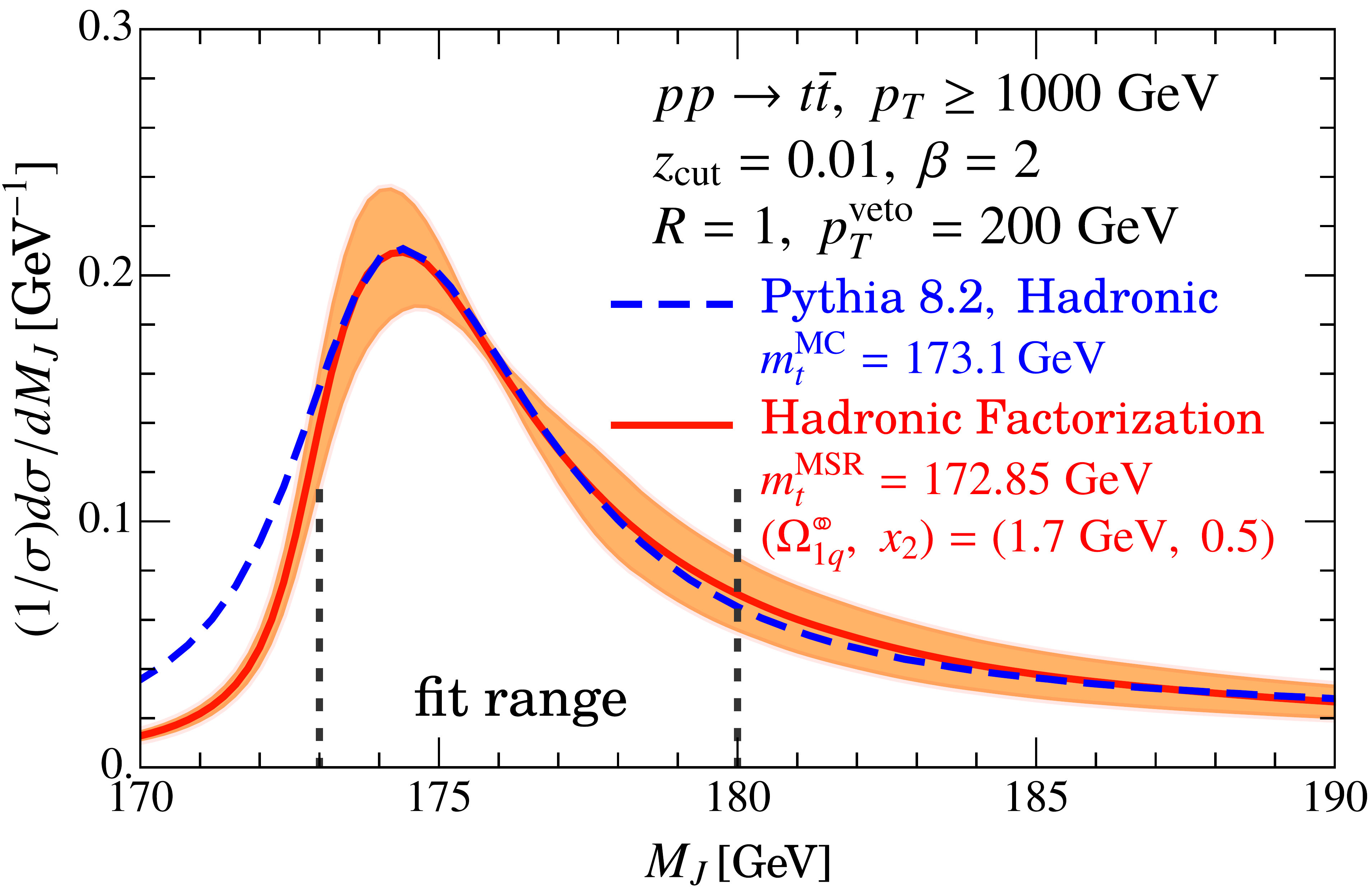}
	 
	\raisebox{4.4cm}{c)} 
	\includegraphics[width=0.44\textwidth]{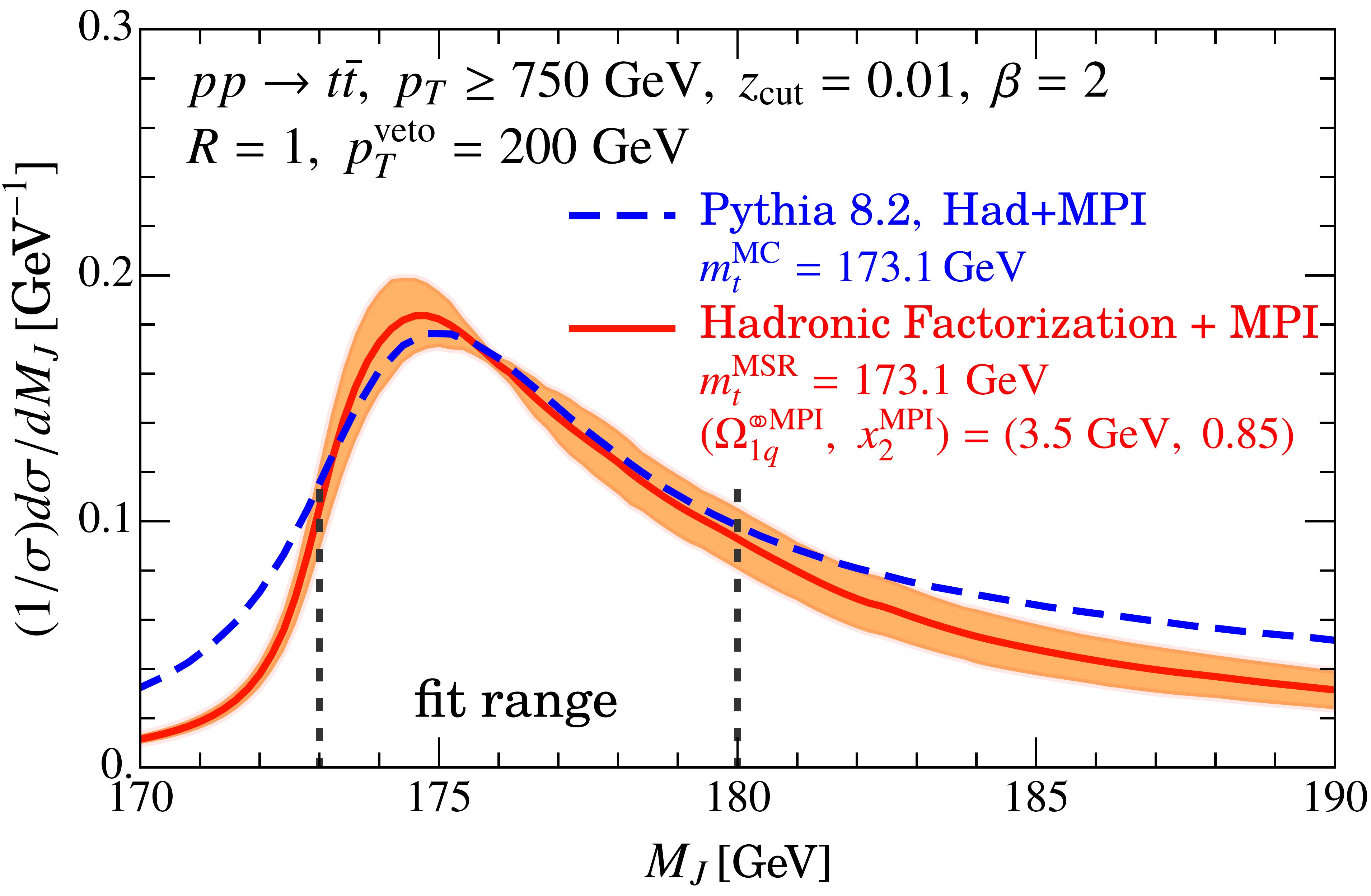}
	 
	\raisebox{4.4cm}{d)}
	\includegraphics[width=0.44\textwidth]{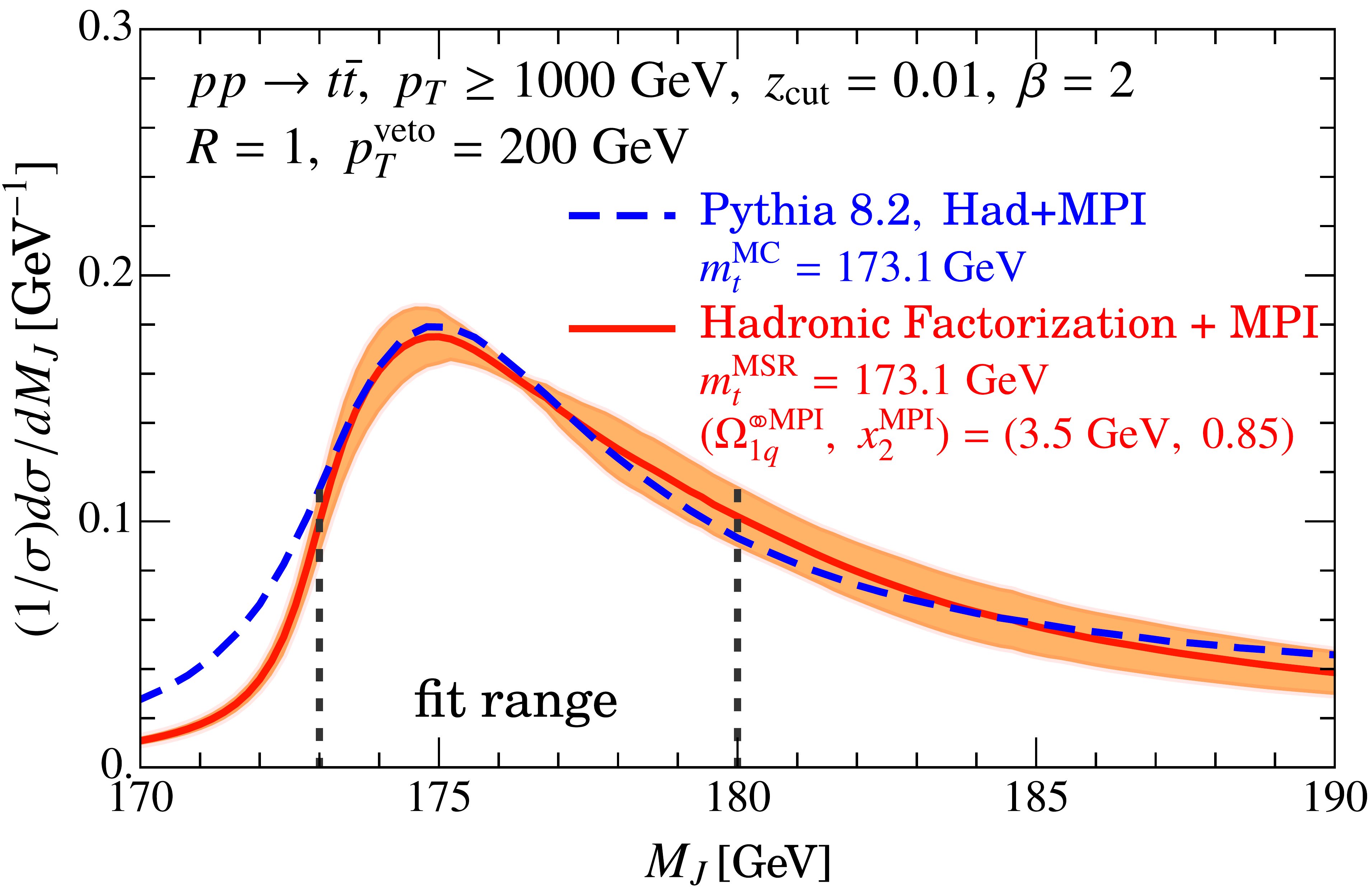}
	 
	\caption{Comparison of \Pythiaeight without and with MPI to the  factorization theorem at NLL with $m_t$ in the MSR mass scheme. \label{fig:pythiatheory} 
	}
	\vspace{-0.2cm}
\end{figure}

In Fig.~\ref{fig:pythiatheory} we show a direct comparison between \Pythiaeight results and the factorization formula in \eq{factthmfinalNLL} at NLL order. For the factorization prediction we use $\alpha_s(m_Z)=0.118$ and the MSR short distance top mass $m_t^{\rm MSR}(R_{m,0}=1~\mbox{GeV})$ as the reference input. 
Furthermore we employ the non-perturbative hadronization parameters
\begin{align}
\label{eq:x2def}
\Omega_{1q}^{\figeight}  \qquad\text{and}\qquad  x_2\equiv \frac{\Omega_{2q}^{\figeight}}{(\Omega_{1q}^{\figeight})^2}-1  \,,
\end{align} 
see \eq{momentsn}, as input to fix the form of the shape function $F_{\figeight}^q(k)$.

In order to extend our hadron level factorization formula to account for the UE we make use of the results of Ref.~\cite{Stewart:2014nna} where it was shown that MPI in Pythia for the ungroomed jet mass spectrum can be well modeled by simply changing parameters in the nonperturbative shape function. This occurs because the dominant impact of MPI is to populate the jet with uncorrelated soft radiation of somewhat higher energy than that associated to the soft hadronization. We adopt this approach to account for hadronization plus UE, replacing 
\begin{align}
\Omega_{nq}^{\figeight} \to \Omega_{nq}^{\figeight{\rm MPI}} \,.
\end{align} 
Estimating that this treatment of UE is uncertain at the $\lesssim 30\%$ level, this induces a residual uncertainty of $\Delta m_t \lesssim 0.3\,{\rm GeV}$ for our soft drop top mass extraction, compared to $\Delta m_t \lesssim 1.4\,{\rm GeV}$ without soft drop. With additional dedicated studies this uncertainty may be further reduced. Lastly, we note that from the work in \Ref{Hoang:2019ceu} we only know about the universality properties of the first moment $\Omega_{1q}^{\figeight}$ of the shape function, whereas the higher moments $n \geq 2$ may depend on the grooming parameters and the kinematic variables in a manner that is not determined solely by $C_1^{q(pp)}$, although its inclusion does capture the proper power counting for these terms. 
Technically, the higher moment $\Omega_{2q}^{\figeight}$ should involve an additional Wilson coefficient which we have not derived. Since $\Omega_{2q}^{\figeight}$ gives a sub-dominant power correction our approximation should be reasonable.

For the \Pythiaeight results we use the Monte-Carlo mass $m_t^{\rm MC}=173.1\,{\rm GeV}$ as input and employ the default Monash 2013 tune~\cite{Skands:2014pea} of the Lund string fragmentation model for its hadronization corrections. To achieve a meaningful comparison of our factorization prediction with the \Pythiaeight results
we carry out a simultaneous fit for $m_t^{\rm MSR}(R_{m,0}=1\,{\rm GeV})$, $\Omega_{1q}^{\figeight}$ and $x_2$ entering the formula in \eq{factthmfinalNLL} to the \Pythiaeight results simultaneously including $p_T\ge 750\,{\rm GeV}$ and $p_T\ge 1000\,{\rm GeV}$ bins. 
For the fit range we take $M_J\in [173,180]\,{\rm GeV}$ and utilize 10 $M_J$ bins in this range, over which the central factorization curve and the \Pythia curve are also normalized. The fits are carried out independently for \Pythiaeight with only hadronization (with the results shown in Figs.~\ref{fig:pythiatheory}a,b), and for \Pythiaeight with both hadronization and MPI (with the results shown in Figs.~\ref{fig:pythiatheory}c,d). 
The resulting best fit values for $m_t^{\rm MSR}(1\,{\rm GeV})$, $\Omega_{1q}^{\figeight}$ 
($\Omega_{nq}^{\figeight{\rm MPI}}$) and $x_2$ ($x_2^{{\rm MPI}}$) are displayed in the panels of Fig.~\ref{fig:pythiatheory}.
For the fits to \Pythiaeight with both hadronization and MPI turned on
we expect from our treatment of MPI effects modified fit results for the non-perturbative parameters 
$\Omega_{1q}^{\figeight{\rm MPI}}$ and $x_2^{{\rm MPI}}$ in comparison to $\Omega_{1q}^{\figeight}$ and $x_2$ obtained without MPI
while the fit result for $m_t^{\rm MSR}(1\,{\rm GeV})$ should remain unchanged.

The jet mass spectra obtained from the factorization theorem and \Pythiaeight, shown in the panels of Fig.~\ref{fig:pythiatheory}, are in quite good agreement for both $p_T$ bins as well as for the different treatments of MPI. As expected from the general structure of the factorization theorem the peak is essentially at the same location for each $p_T$ bin. However, there is a noticeable difference between the factorization theorem results and \Pythiaeight for the tail on the left of the peak. This is related to the grooming of radiation that is emitted off the primary top decay products that ends up outside of the groomed jet and thus enhances the spectrum to the left of the peak. 
While \Pythiaeight simulates this final state radiation, it is absent from our current decay function at the level it has been formulated here. The curves in 
Figs.~\ref{fig:pythiatheory}a,b,c,d show that this final state radiation does not affect significantly the distribution in the peak region, indicated by our fit range.

\begin{figure}[t!]
	\includegraphics[width=0.44\textwidth]{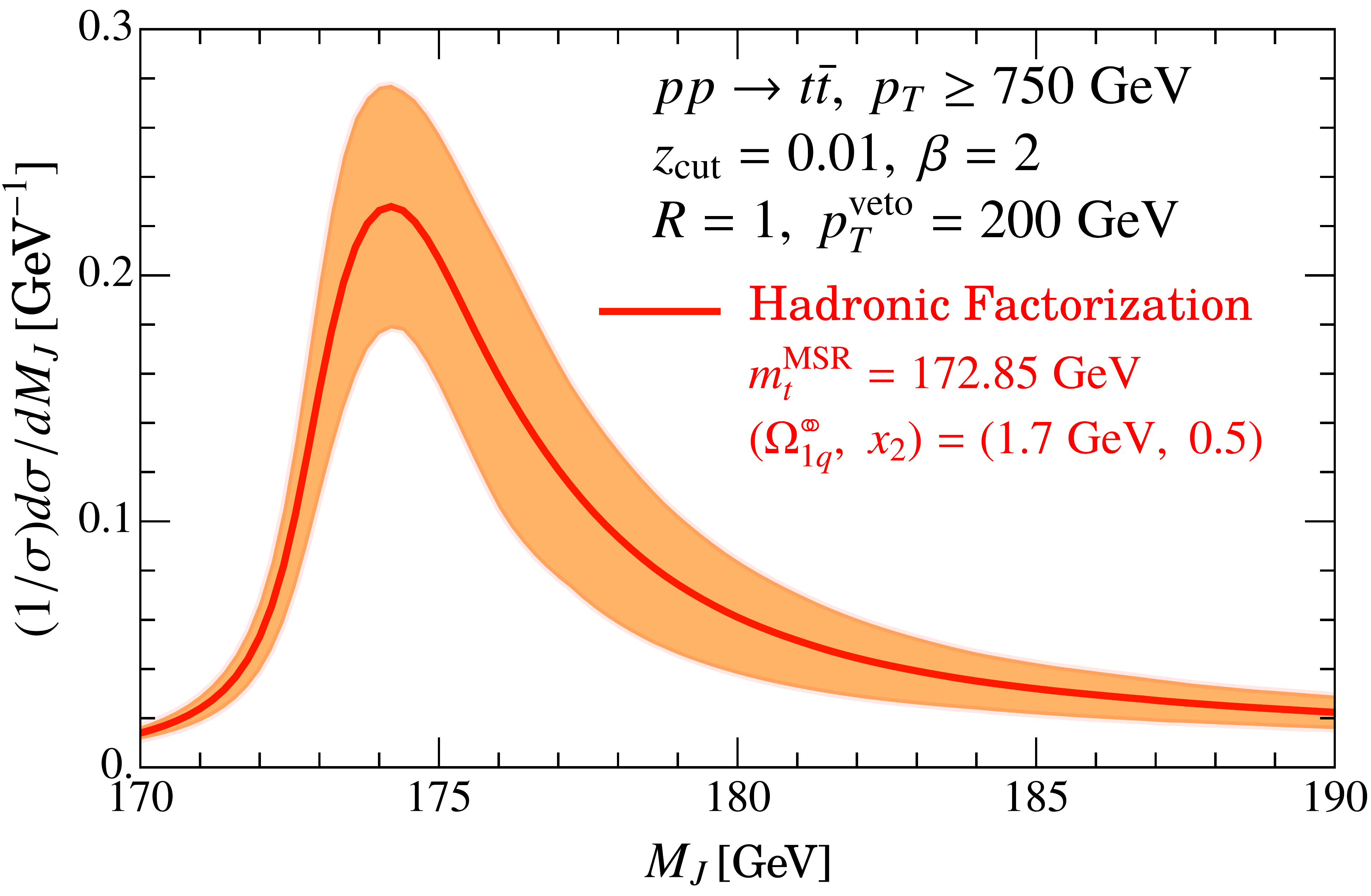}
	\vspace{-0.2cm}
	 
	\caption{Perturbative NLL uncertainty in the cross section without normalizing the scale variations. 
		\label{fig:normUncert} 
	}
	\vspace{-0.2cm}
\end{figure}

In the panels of Figs.~\ref{fig:pythiatheory} we also show as the orange bands the theoretical uncertainty of our NLL factorization formula varying renormalization scales. The bands are the envelope of the varying curves each of which is normalized to the fit range so that only the uncertainties on the shape are displayed. In contrast, in \fig{normUncert} the orange uncertainty band include the uncertainties on the normalization. Here we normalize the central curve as before in the range $M_J \in [173, 180]$~GeV, and use this same normalization for the cross sections obtained for the varied renormalization scales. Not  unexpectedly, the normalization of the cross section is rather poorly determined by results at NLL order. Thus, from comparing \figs{pythiatheory}{normUncert} we conclude that the normalized shape is more robust against higher order corrections, and hence should be used in comparing with data or MC results. 
However, we remind the reader that here we do not carry out a complete analysis of uncertainties for our fit at NLL, and hence have not assigned uncertainties to our fit parameter results. For the fits done here, only the central theory curves were used.
From the size of the NLL uncertainty bands, and accounting for MPI modeling uncertainty, we estimate the precision of this analysis  concerning the sensitivity to the top quark mass to be at the level of 1 GeV. 
This is dominated by perturbative uncertainties that are expected to significantly decrease when increasing our analysis from NLL to next-to-next-to-leading logarithmic order (NNLL), at which point a complete analysis of theoretical uncertainties becomes warranted. 

With these caveats about perturbative uncertainties in mind, the results shown in Figs.~\ref{fig:pythiatheory} still allow to make a number of instructive observations concerning the interpretation of the \Pythiaeight top mass parameter $m_t^{\rm MC}$.
The fit values for $m_t^{\rm MSR}(1~\mbox{GeV})$  obtained in  Figs.~\ref{fig:pythiatheory}a,b,c,d are all within $0.3\, {\rm GeV}$ of the input $m_t^{\rm MC}$. This result is compatible with the $e^+e^-$ calibration result in~\cite{Butenschoen:2016lpz} for the ungroomed 2-jettiness distribution in the peak region and
supports universality of top mass MC calibration fits for $pp$ groomed and $e^+e^-$ ungroomed jet masses even with our model treatment of MPI effects. 

We also observe that the fit values of $m_t^{\rm MSR}(1~\mbox{GeV})$  agree within $0.3\,{\rm GeV}$ for the analyses with and without MPI effects. This fully confirms our anticipation that the dominant effect of adding MPI is an increase of  the scale of the hadronization parameter $\Omega_1^{\figeight}\to \Omega_1^{\figeight{\rm MPI}}$ and to modify $x_2\to x_2^{{\rm MPI}}$, while the top quark mass remains unchanged.  This is a crucial outcome and supports that a precision determination of the top quark mass  may be achieved from boosted top jet measurements.
It furthermore supports that MC generator top mass calibration analyses along the lines of~\Ref{Butenschoen:2016lpz} can be carried out successfully in the LHC environment using groomed boosted top quark initiated jets.
However, we also point out that when examining the lower $p_T$ bin $[550,750]\,{\rm GeV}$ we find poorer agreement with \Pythiaeight, which likely indicates that higher order terms in the soft drop factorization expansions are becoming important for smaller boosts.

In the \App{pole} we present the corresponding fit results for the pole mass scheme, obtaining values that are about $0.6$ GeV smaller than $m_t^{\rm MC}$, which implies that our $pp$ groomed pole mass analysis is also compatible with the $e^+e^-$ ungroomed pole mass calibration study in~\cite{Butenschoen:2016lpz}. 

\section{Conclusion}

In this paper we have derived a factorization formula for soft drop groomed jets involving a boosted top quark, which can be used to directly access short distance top masses in a hadronic collider, while avoiding significant contamination from soft radiation.
Since our approach is systematically improvable, we anticipate that the perturbative and hadronization uncertainties on extracting $m_t$ from soft drop groomed top initiated jet mass distributions will eventually be below a ${\rm GeV}$.   Experimentally requiring $p_T$ above $700\,{\rm GeV}$ limits the data sample, however this is somewhat mitigated by the light soft drop method not requiring other cuts, like those on $p_T^{\rm veto}$ or the angle between decay products used in Ref.~\cite{Sirunyan:2017yar}. We leave a more detailed analysis to future work, including results at one higher order, more precise estimates of all uncertainties, and exploring the smaller $p_T$ region. We anticipate that these results can be used for direct fits to LHC data, to calibrate the MC top mass parameter, and even to make predictions independent of fitting hadronic parameters by exploiting the fact that universality allows $\Omega_{1q}^{\figeight{\rm MPI}}$ to be determined by fits to light or $b$ quark soft dropped jet data. We also leave further exploration of the residual dependence on $x_2$ and higher moments to future studies~\cite{Hoang:2018SDlong}. 

It would also be interesting to further test fits to the $M_J$ spectrum including pileup interactions and using a range of the commonly used pileup mitigation techniques, including the use of the light soft drop proposed here or other methods that avoid disrupting the top decay products. We anticipate that after this mitigation any remaining radiation from pileup will again behave similar to the underlying event, adding uncorrelated radiation to the soft drop jet region, and hence will simply further modify the fit values of our $\Omega_{1q}^{\figeight{\rm MPI}}$ and $x_2^{{\rm MPI}}$ parameters. It will be interesting to test this with a dedicated study in the future. 

\vspace{0.2cm}
{\bf Acknowledgment:} 
We thank Marcel Vos, Roman Kogler, Jesse Thaler, and Andrew Larkoski for discussions. We acknowledge partial support from the FWF Austrian Science Fund under the Project No.~P28535-N27, the U.S. Department of Energy under Grant No.~DE-SC0011090, and the Simons Foundation through Grant No.~327942. We also thank the Erwin-Schr\"odinger International Institute for Mathematics and Physics, the University of Vienna and Cultural Section of the City of Vienna (MA7) for partial support.

\vspace{0.2cm}

\appendix

\section{Wilson Coefficient for Shift Power Correction}
\label{NPcoeff}
Here we state results from \Ref{Hoang:2019ceu} for the nonperturbative Wilson Coefficient  $C_1^{q(pp)}(m_J^2)$ introduced in \eq{factfull}. The calculation at LL is most intuitively carried out in the coherent branching formalism where the resummation is organized as a sum over independent real emissions, allowing one to keep track of the branches that pass and fail soft drop. The result for $C_1^{q(pp)}(m_J^2)$ is then given by
\begin{align}  \label{eq:C1pp}
 & C_1^{q(pp)}(m_J^2, Q, \tilde z_{\rm cut}, \beta, R) 
 \nn\\
 &= \frac{R}{\cosh \eta_J} \, \tilde C_1^{q(pp)} \big (m_J^2,  Rp_T, z_{\rm cut}'R^\beta,  \beta\big) 
\, ,
\end{align}
where
\begin{align}
\label{eq:C1}
& \tilde C_1^{q(pp)} \big (m_J^2,  Rp_T, z_{\rm cut}'R^\beta,  \beta\big) =  \frac{e^{-{\cal R}_q\bigl(\frac{m_J^2}{R^2p_T^2}\bigr) }  }{C_0^{q(pp)}(m_J^2)} \nn \\
& \qquad \times 
\int_{0}^{1} \frac{d\tilde \theta^2}{\tilde \theta^2}\,\frac{\tilde \theta}{2}\alpha_s\Big(\frac{m_J^2}{\tilde \theta \, Rp_T} \Big)
\frac{C_F}{\pi} \frac{m_J^2}{\tilde \theta^2 R^2p_T^2}  
p_{gq}\Big(\frac{m_J^2}{\tilde \theta^2 R^2p_T^2} \Big) \nn \\
&\qquad \qquad \times
\Theta \Bigl(\tilde \theta -\frac{m_J}{Rp_T} \Bigr) \,
\Theta\big(\tilde \theta^\star(m_J^2) - \tilde \theta \big) \, ,
\end{align}
with
\begin{align}
\label{eq:zcutprime}
z'_{\rm cut} = z_{\rm cut}/R_0^\beta \, .
\end{align} 
Here $\tilde \theta^\star = \theta^\star/R$ denotes the maximum angle of the soft drop stopping subjet at a given jet mass $m_J^2$, and is given by 
\begin{align}
\label{eq:thetastar}
\tilde \theta^\star(m_J^2) &= 
\tilde \theta^\star(m_J^2, Rp_T, z'_{\rm cut} R^\beta, \beta ) \\
&
\qquad \equiv\frac{2\cosh\eta_J}{R}\,\bigg(\frac{ m_J^2}{ Q \qcut}\bigg)^{\frac{1}{2 + \beta}}  \,. \nn
\end{align}
The splitting function $p_{gq}(z)$ has the form
\begin{align}
p_{gq}(z) = \frac{1 + (1-z)^2}{2z} \, .
\end{align}
Finally, the prefactor $e^{-{\cal R}_q}/C_0^{q(pp)}$ in \eq{C1} is given by
\begin{align} 
\label{eq:C0}
&
\frac{e^{-{\cal R}_q\bigl(\frac{m_J^2}{R^2p_T^2}\bigr) }}{ C_0^{q(pp)}\big (m_J^2\big)}
=\frac{e^{-{\cal R}_q\bigl(\frac{m_J^2}{R^2p_T^2},Rp_T, z'_{\rm cut} R^\beta,\beta\bigr) }}{ C_0^{q(pp)}\big (m_J^2,Rp_T, z'_{\rm cut} R^\beta,\beta\big)} 
\\
&
\qquad
=\bigg[ \int_{0}^{1} \frac{d\tilde \theta^2}{\tilde \theta^2}\: \alpha_s\Big(\frac{m_J^2}{\tilde \theta Rp_T} \Big)\frac{C_F}{\pi} \, \frac{m_J^2}{\tilde \theta^2R^2 p_T^2 }  \: p_{gq}\Big(\frac{m_J^2}{\tilde \theta^2R^2 p_T^2 }\Big)\nn \\
&\qquad\qquad \times \Theta \Bigl(\tilde \theta -\frac{m_J}{Rp_T} \Bigr) \, \Theta\Bigl(\tilde \theta_{pp}^\star(m_J^2) - \tilde \theta \Bigr) \bigg]^{-1}
 \, . \nn
\end{align}

The result for $C_1(m_J^2)$ is valid in the soft drop operator expansion region~\cite{Hoang:2019ceu} given by combination of \eqs{sdlimit}{sdoe}:
\begin{align}
\label{eq:SDOE}
\frac{Q \Lambda_{\rm QCD}}{m_J^2}
\Bigl( \frac{m_J^2}{Q Q_{\rm cut}}\Bigr)^{\frac{1}{2+\beta}}  
\ll 1 \, , \quad m_J^2 \ll  z_{\rm cut}  \frac{Q^2}{4} \, .
\end{align}
The first inequality in \eq{SDOE} guarantees that the collinear soft radiation is perturbative and the second one ensures that soft drop is active. Replacing the ``$a\gg b$'' by ``$a=3 b$'' and $m_J^2$ by $m_t \hat s_t$ in \eq{SDOE} we find that for top quark jets the constraints are satisfied for $M_J \in [180,187]$ GeV for $p_T \geq 750$ GeV, $|\eta_{t,\tb} |< 2.5$ and our standard choice of soft drop parameters, $\zcut = 0.01$ and $\beta = 2$. From \fig{thcomp} we find that this also happens to be precisely the region where the high-$p_T$ factorization starts to become comparable to the decay factorization. In the peak region for $M_J \lesssim 180$ GeV the decay products predominantly stop soft drop and $C_1$ has little role to play. For the far tail region $M_J \gtrsim 190$ GeV one needs to include fixed order corrections to account for transition to the no-soft-drop region, which we leave to future studies.

\begin{figure}[t!]
	\raisebox{4.4cm}{a)}
	\includegraphics[width=0.44\textwidth]{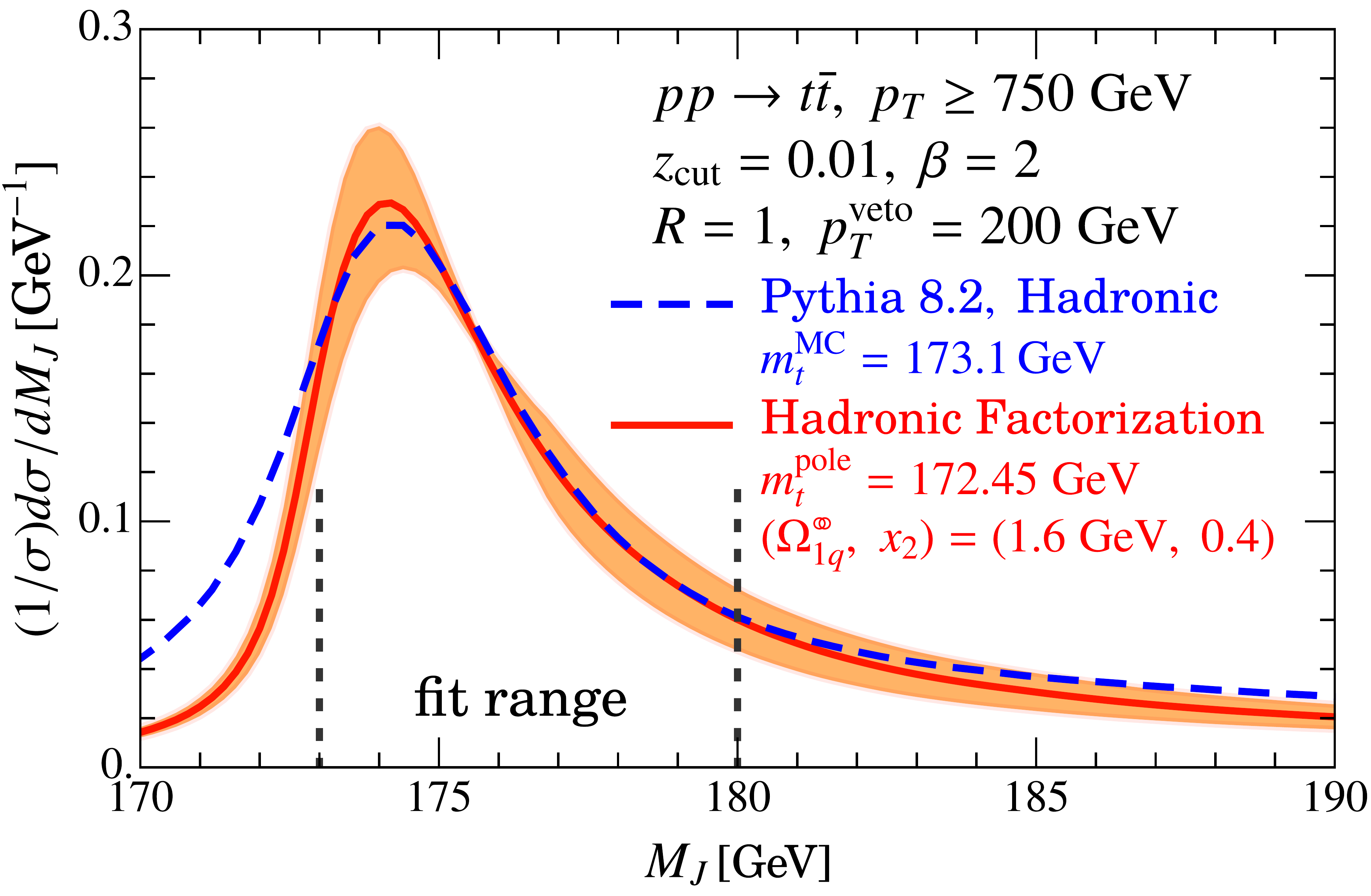}
	 
	\raisebox{4.4cm}{b)}
	\includegraphics[width=0.44\textwidth]{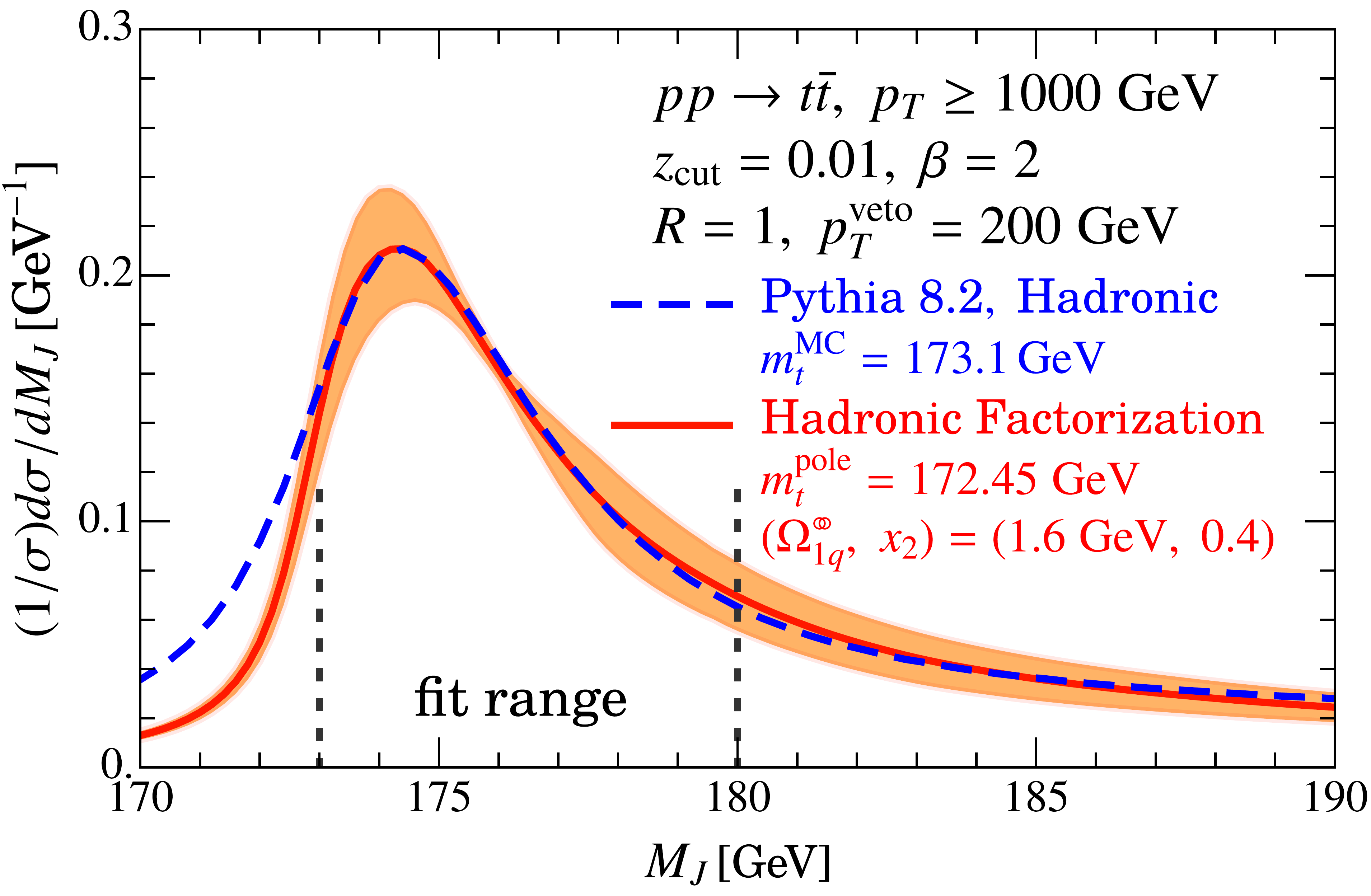}
	 
	\raisebox{4.4cm}{c)} 
	\includegraphics[width=0.44\textwidth]{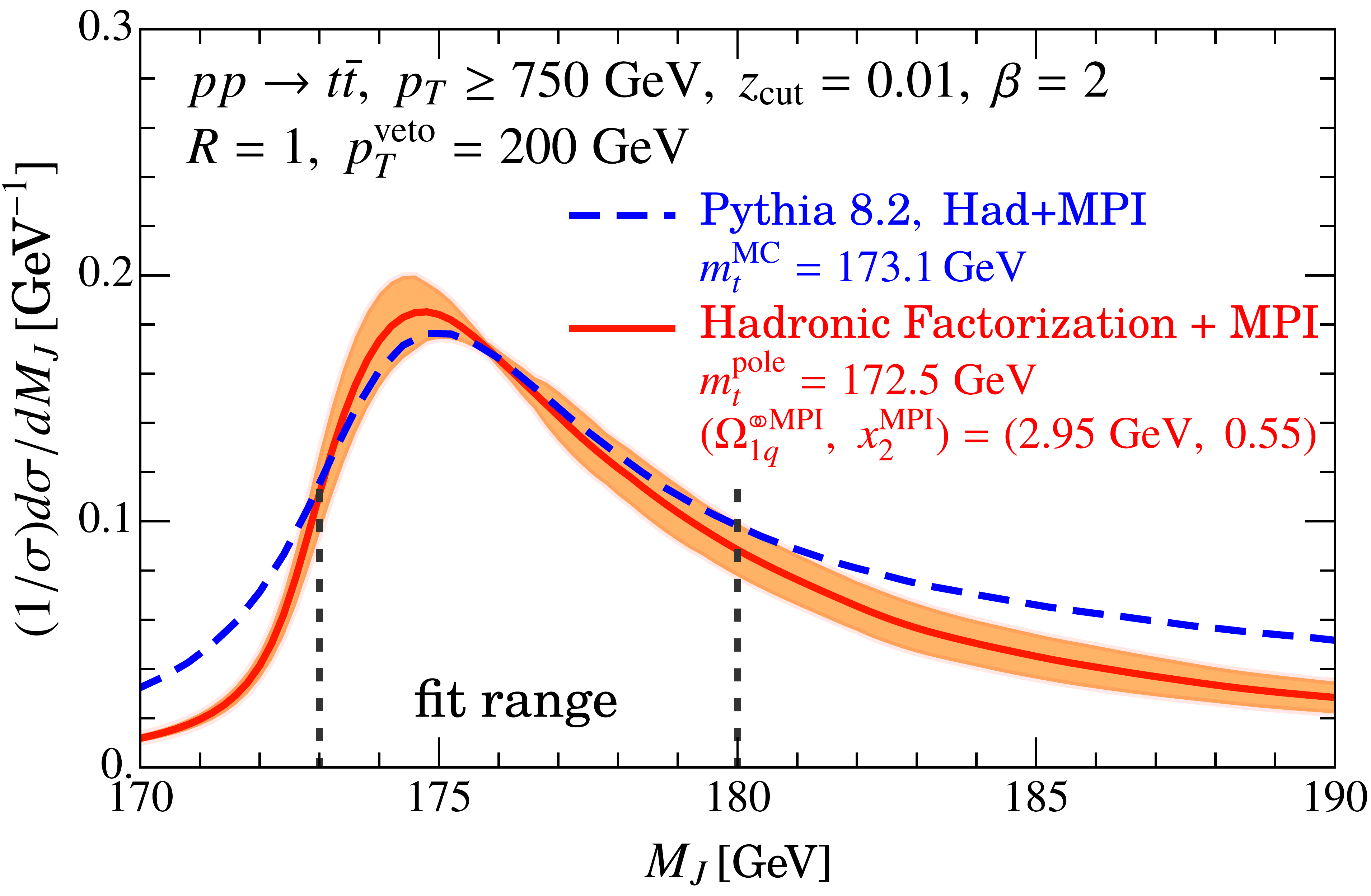}
	 
	\raisebox{4.4cm}{d)}
	\includegraphics[width=0.44\textwidth]{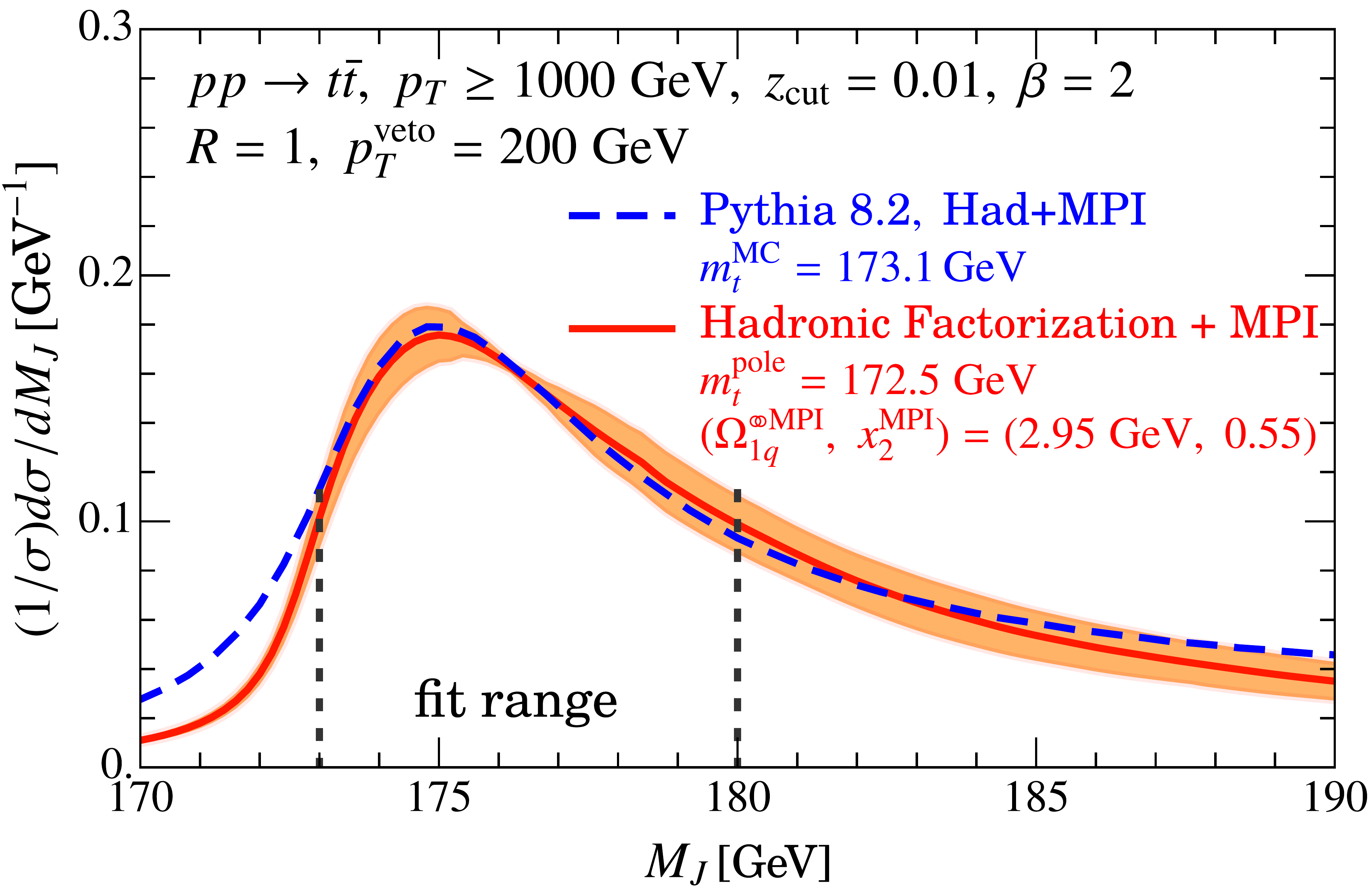}
	 
	\caption{Comparison of \Pythiaeight without and with MPI to the  factorization theorems at NLL with $m_t$ in the pole-mass scheme. \label{fig:pythiatheorypole} 
	}
	\vspace{-0.2cm}
\end{figure} 

\section{Results in the pole mass scheme}\label{pole}

In this appendix we present results for fits of our factorization formula in \eq{factthmfinalNLL}
to \Pythiaeight with hadronization and with hadronization+MPI 
employing the pole mass scheme $m_t^{\rm pole}$. 
In  Figs.~\ref{fig:pythiatheorypole}a-d the pole mass scheme analogues of the results in
Figs.~\ref{fig:pythiatheory}a-d for the MSR mass scheme are shown. Apart from the mass scheme change, we have carried out identical analyses, and we again find that the spectra obtained from the factorization theorem and \Pythiaeight are in quite good agreement.

We observe that the fitted $m_t^{\rm pole}$ values are around $0.6$~GeV smaller than the input $m_t^{\rm MC}$. 
This is again compatible with the corresponding $e^+e^-$ calibration result in~\Ref{Butenschoen:2016lpz} for the ungroomed 2-jettiness distribution. The results supports the conclusion of~\Ref{Butenschoen:2016lpz} that the pole mass cannot be directly identified with the MC top mass parameter $m_t^{\rm MC}$. As for our MSR mass analysis, the obtained fit values of $m_t^{\rm pole}$ are also compatible within uncertainties for fits to \Pythiaeight 
with and without MPI effects.  As for the fits in the MSR scheme, the dominant effect of adding MPI is to significantly increase the scale of the hadronization parameter $\Omega_1^{\figeight}\to \Omega_1^{\figeight{\rm MPI}}$ and to modify $x_2\to x_2^{{\rm MPI}}$. 
Interestingly, the values obtained for the $\Omega_1^{\figeight}$ and $x_2$ parameters with the top mass in the pole mass scheme are very similar to those of our MSR mass analysis while the values for  $\Omega_1^{\figeight{\rm MPI}}$ and $x_2^{\rm MPI}$ slightly differ. At the level of the precision of our analysis, the difference is, however, not significant.

We note that the reason the fitted top mass values in the pole mass scheme are $0.4$--$0.6\,{\rm GeV}$ smaller than the corresponding $m_t^{\rm MSR}(1\,{\rm GeV})$ fit results can be analytically tracked down to the way both schemes have been implemented at NLL order (see the last paragraph of \sec{topfact}).
At NLL order the only difference in the factorization formulae for the two mass schemes comes from the evolution of $m_t^{\rm MSR}(R_m)$ where $R_m$ is set to the scale of the  bHQET jet function $J_B$. 
Since the typical scale of $J_B(\hat s_t,\delta m,\Gamma_t,\mu)$ is
around $5$~GeV in the peak region of the jet mass distribution, which provides the highest top mass sensitivity, it is effectively $m_t^{\rm MSR}(R_m\simeq 5\,\mbox{GeV})$ that is constrained by the fitting procedure. Thus at NLL order the fitted value for $m_t^{\rm MSR}(R_m\simeq 5\,\mbox{GeV})$ in the MSR mass analysis is very close to the fitted value of $m_t^{\rm pole}$ in the pole mass analysis. The numerical difference between the obtained fit results for $m_t^{\rm MSR}(1\,{\rm GeV})$ and
$m_t^{\rm pole}$ therefore predominantly arises from the $R$-evolution between $\sim 5$~GeV and
the reference scale $R_{m,0}=1$~GeV,  $m_t^{\rm MSR}(1\,{\rm GeV})= m_t^{\rm MSR}(5\,{\rm GeV}) + 0.53\,{\rm GeV}$.

It is known that the pole mass has a renormalon ambiguity of $\sim \Lambda_{\rm QCD}$, thus the determination of $m_t^{\rm pole}$ at higher order can in general expected to be more uncertain than that of the short-distance MSR mass. This is compatible with interpreting, from the point of view of anticipated higher order analyses, the difference between the results from directly fitting for $m_t^{\rm pole}$, and obtaining $m_t^{\rm pole}$ via the MSR fit result via  $m_t^{\rm pole}= m_t^{\rm MSR}(1\,{\rm GeV}) +0.17\,{\rm GeV}$, as an additional uncertainity in the pole mass.

However, we note that when considering only results at NLL order, the numerical difference between the fitted pole and MSR mass values can be attributed to the fact that the pole mass scheme treats real radiation at all scales as resolved, while the MSR mass $m_t^{\rm MSR}(R_m)$ treats real radiation for scales below the scale $R_m$ as unresolved. The conclusion that this leads to an ambiguity in the pole mass definition only arises in the context of considering higher order perturbative corrections.

\vfill

\bibliographystyle{apsrev4-1}

\bibliography{../top3}

\newpage
\clearpage

\end{document}